\documentclass[12pt]{article}
\usepackage{graphicx}
\usepackage{a4p}
\usepackage{amssymb}
\def\sqeeb{\ifmmode{\sqrt{s_{\protect\bf\mathrm{ee}}}}\else
  {$\sqrt{s_{\protect\bf\mathrm{ee}}}$}\fi}
\def\epem{\ifmmode{\mathrm{e}^{+}\mathrm{e}^{-}}\else
  {$\mathrm{e}^{+}\mathrm{e}^{-}$}\fi}
\def\sqee{\ifmmode{\sqrt{s_\mathrm{ee}}}\else
  {$\sqrt{s_\mathrm{ee}}$}\fi}
\def\kp{{\ifmmode{k_{\perp}}\else{$k_{\perp}$}\fi}}
\def\etmean{\ifmmode{\bar{E}^{\mathrm{jet}}_{\mathrm{T}}}\else
  {$\bar{E}^{\mathrm{jet}}_{\mathrm{T}}$}\fi}
\def\etajmean{\ifmmode{|\bar{\eta}^\mathrm{jet}|}\else 
  {$|\bar{\eta}^\mathrm{jet}|$}\fi}
\def\lxg{\ifmmode{\mathrm{log_{10}}(x_{\gamma})}\else
  {${\mathrm{log_{10}}(x_{\gamma})}$}\fi}
\def\xg{\ifmmode{x_{\gamma}}\else{${x_{\gamma}}$}\fi}
\def\xgp{\ifmmode{x_{\gamma}^+}\else{${x_{\gamma}^+}$}\fi}
\def\xgm{\ifmmode{x_{\gamma}^-}\else{${x_{\gamma}^-}$}\fi}
\def\xgpm{\ifmmode{x_{\gamma}^{\pm}}\else 
  {${x_{\gamma}^{\pm}}$}\fi}
\def\ipb{\ifmmode {\mathrm{pb}^{-1}}\else 
  {$\mathrm{pb}^{-1}$}\fi}
\def\as{\alpha_{\rm s}}
\def\ee{\ifmmode{\mbox{e}^+\mbox{e}^-}\else
  {$\mbox{e}^+\mbox{e}^-$}\fi}
\def\thetamaxp{\ifmmode{\theta_\mathrm{max}'}\else
  {$\theta_\mathrm{max}'$}\fi}
\def\pt{\ifmmode{p_\mathrm{T}}\else{$p_\mathrm{T}$}\fi}
\def\Zzero{\ifmmode{\mathrm{Z}^{0}}\else{$\mathrm{Z}^{0}$}\fi}

\def\etjet{\ifmmode{E^\mathrm{jet}_\mathrm{T}}\else 
  {$E^\mathrm{jet}_\mathrm{T}$}\fi}
\def\ptmiss{\ifmmode{{P}_{\mathrm{T,MISS}}}\else 
  {${P}_{\mathrm{T,MISS}}$}\fi}
\def\mj1h2{\ifmmode{M_{\mathrm{J1H2}}}\else 
  {$M_{\mathrm{J1H2}}$}\fi}
\def\ebeam{\ifmmode{E_{\mathrm{BEAM}}}\else 
  {$E_{\mathrm{BEAM}}$}\fi}
\def\etajet{\ifmmode{|\eta^\mathrm{jet}|}\else 
  {$|\eta^\mathrm{jet}|$}\fi}
\def\etaj{\ifmmode{\eta^\mathrm{jet}}\else 
  {$\eta^\mathrm{jet}$}\fi}
\def\etajdef{\ifmmode{\eta^\mathrm{jet} = 
    -\ln\tan(\theta^\mathrm{jet}/2)}\else{$\eta^\mathrm{jet} = 
    -\ln\tan(\theta^\mathrm{jet}/2)$}\fi}
\def\detajet{\ifmmode{|\Delta\eta^\mathrm{jet}|}\else 
  {$|\Delta\eta^\mathrm{jet}|$}\fi}
\def\costhst{\ifmmode{|\mathrm{cos}\,\Theta^{*}|}\else 
  {$|\mathrm{cos}\,\Theta^{*}|$}\fi}
\def\etajetc{\ifmmode{|\eta^\mathrm{jet}_\mathrm{cntr}|}\else 
  {$|\eta^\mathrm{jet}_\mathrm{cntr}|$}\fi}
\def\etajetf{\ifmmode{|\eta^\mathrm{jet}_\mathrm{fwd}|}\else 
  {$|\eta^\mathrm{jet}_\mathrm{fwd}|$}\fi}
\def\etah{\ifmmode {\hat{\eta}}\else{$\hat{\eta}$}\fi}
\def\dsdetm{\frac{\mathrm{d}{\sigma}_{\mathrm{dijet}}}
  {\mathrm{d}\bar{E}^\mathrm{jet}_\mathrm{T}}}

\def\dsdxg{\frac{\mathrm{d}{\sigma}_{\mathrm{dijet}}}
  {\mathrm{d}x_{\gamma}}}
\def\dsdlxg{\frac{\mathrm{d}{\sigma}_{\mathrm{dijet}}}
  {\mathrm{d log_{10}}\left(x_{\gamma}\right)}}
\def\dsdeta{\ifmmode{\frac{\mathrm{d}{\sigma}_{\mathrm{dijet}}}
  {\mathrm{d}\etajet}}\else
    {$\frac{\mathrm{d}{\sigma}_{\mathrm{dijet}}}
      {\mathrm{d}\etajet}$}\fi}
\def\dsdetac{\frac{\mathrm{d}{\sigma}_{\mathrm{dijet}}}
  {\mathrm{d}\etajetc}}
\def\dsdetaf{\frac{\mathrm{d}{\sigma}_{\mathrm{dijet}}}
  {\mathrm{d}\etajetf}}
\def\dsddeta{\frac{\mathrm{d}{\sigma}_{\mathrm{dijet}}}
  {\mathrm{d}\detajet}}
\def\dscost{\frac{\mathrm{d}{\sigma}_{\mathrm{dijet}}}
  {\mathrm{d}\costhst}}
\def\et{\ifmmode{E_\mathrm{T}}\else{$E_\mathrm{T}$}\fi}

\def\gg{\ifmmode{\gamma\gamma}\else{$\gamma\gamma$}\fi}
\def\gsg{\ifmmode{\gamma^{\star}\gamma}\else
  {$\gamma^{\star}\gamma$}\fi}
\def\PTMIA{\ifmmode{p_\mathrm{t}^\mathrm{mi}}\else
  {$p_\mathrm{t}^\mathrm{mi}$}\fi}
\def\sas1d{SaS\,1D}

\def\grv{GRV}
\def\grvnlo{GRV\,HO}
\def\afgnlo{AFG\,HO}
\def\gs96nlo{GS96\,HO}
\def\lac1{LAC\,1}
\def\hadcor{\ifmmode{(1+\delta_{hadr})}\else{$(1+\delta_{hadr})$}\fi}
\begin{document}
\begin{titlepage}

\begin{center}
{\large EUROPEAN ORGANIZATION FOR NUCLEAR RESEARCH}
\end{center}
\vspace*{1.0cm}

\begin{flushright}
CERN-EP-2002-093\\
20 December 2002
\end{flushright}
\vspace*{1.0cm}

\begin{center}
{\boldmath\LARGE\bf Di-Jet Production in Photon-Photon \\
\vspace*{0.3cm}
Collisions at $\sqeeb$ from $189$ to $209$~GeV \unboldmath}
\end{center}
\vspace*{0.5cm}

\begin{center} {\LARGE The OPAL Collaboration} \end{center}
\vspace*{0.5cm}

\begin{center}
{\large  Abstract}
\end{center}
Di-jet production is studied in collisions of quasi-real photons at
{\epem} centre-of-mass energies {\sqee} from 189 to 209~GeV at LEP.
The data were collected with the OPAL detector.  Jets are
reconstructed using an inclusive {\kp}-clustering algorithm for all
cross-section measurements presented.  A cone jet algorithm is used in
addition to study the different structure of the jets resulting from
either of the algorithms.  The inclusive di-jet cross-section is
measured as a function of the mean transverse energy {\etmean} of the
two leading jets, and as a function of the estimated fraction of the
photon momentum carried by the parton entering the hard sub-process,
{\xg}, for different regions of {\etmean}.  Angular distributions in
di-jet events are measured and used to demonstrate the dominance of
quark and gluon initiated processes in different regions of phase
space.  Furthermore the inclusive di-jet cross-section as a function
of {\etajet} and {\detajet} is presented, where {\etaj} is the jet
pseudo-rapidity.  Different regions of the {\xgp}-{\xgm}-space are
explored to study and control the influence of an underlying event.
The results are compared to next-to-leading order perturbative QCD
calculations and to the predictions of the leading order Monte Carlo
generator PYTHIA.

\vspace{1.5cm}

\begin{center}
{\sc \large To be submitted to Eur. Phys. J. }
\end{center}

\vfill

\end{titlepage}

\begin{center}{\Large        The OPAL Collaboration
}\end{center}\bigskip
\begin{center}{
G.\thinspace Abbiendi$^{  2}$,
C.\thinspace Ainsley$^{  5}$,
P.F.\thinspace {\AA}kesson$^{  3}$,
G.\thinspace Alexander$^{ 22}$,
J.\thinspace Allison$^{ 16}$,
P.\thinspace Amaral$^{  9}$, 
G.\thinspace Anagnostou$^{  1}$,
K.J.\thinspace Anderson$^{  9}$,
S.\thinspace Arcelli$^{  2}$,
S.\thinspace Asai$^{ 23}$
% ...
}\end{center}\bigskip
\bigskip
$^{  1}$School of Physics and Astronomy, University of Birmingham,
Birmingham B15 2TT, UK
\newline
$^{  2}$Dipartimento di Fisica dell' Universit\`a di Bologna and INFN,
I-40126 Bologna, Italy
\newline
$^{  3}$Physikalisches Institut, Universit\"at Bonn,
D-53115 Bonn, Germany
\newline
% ...
$^{  5}$Cavendish Laboratory, Cambridge CB3 0HE, UK
\newline
% ...
$^{  9}$Enrico Fermi Institute and Department of Physics,
University of Chicago, Chicago IL 60637, USA
\newline
% ...
$^{ 16}$Department of Physics, Schuster Laboratory, The University,
Manchester M13 9PL, UK
\newline
% ...
$^{ 22}$Department of Physics and Astronomy, Tel Aviv University,
Tel Aviv 69978, Israel
\newline
$^{ 23}$International Centre for Elementary Particle Physics and
Department of Physics, University of Tokyo, Tokyo 113-0033, and
Kobe University, Kobe 657-8501, Japan
\newline
% ...
\newpage

\section{Introduction}
\label{sec-intro}
We have studied the production of di-jets in the collisions of two
quasi-real photons at an {\epem} centre-of-mass energy {\sqee} from
189 to 209~GeV, with a total integrated luminosity of \mbox{593~\ipb}
collected by the OPAL detector at LEP.  Di-jet events are of
particular interest, as the two jets can be used to estimate the
fraction of the photon momentum participating in the hard interaction,
which is a sensitive probe of the structure of the photon. The
transverse energy of the jets provides a hard scale that allows such
processes to be calculated in perturbative QCD.  Fixed order
calculations at next-to-leading order (NLO) in the strong coupling
constant $\as$ for di-jet production are available and are compared to
the data, providing tests of the theory.  Leading order Monte Carlo
(MC) generators are used to estimate the importance of soft processes
not included in the NLO calculation.

Inclusive jet cross-sections in photon-photon collisions have
previously been measured at $\sqee=$ 58 GeV at
TRISTAN~\cite{bib-amy,bib-topaz} and at {\sqee} from 130 to 172 GeV at
LEP~\cite{bib-opalgg,bib-djopold}. This paper extends the latter
analysis to higher {\epem} centre-of-mass energies, and provides an
approximately thirty-fold increase in integrated luminosity. The
{\kp}-clustering algorithm~{\cite{bib-ktclus}} is used as opposed to
the cone algorithm~{\cite{bib-coneopal}}
in~\cite{bib-opalgg,bib-djopold} for the measurement of the
differential cross-sections, because of the advantages of this
algorithm in comparing to theoretical
calculations~{\cite{bib-ktisbest}}. The cone jet algorithm is used to
demonstrate the different structure of the cone jets compared to
jets defined by the {\kp}-clustering algorithm.  The large amount of
data allows us to measure the cross-section for di-jet production in
photon-photon interactions as a function of the mean transverse jet
energy {\etmean}, the jet pseudo-rapidity {\etajet} and the absolute
difference in pseudo-rapidity {\detajet} of the jets, with {\etajdef}
\footnote{The coordinate system of OPAL has the $z$ axis along the
  electron beam direction, the $y$ axis pointing upwards and $x$
  towards the center of the LEP ring. The polar angle $\theta$ and the
  azimuthal angle $\phi$ are defined relative to the $+z$-axis and
  $+x$-axis, respectively.}.
For the first time, the differential
cross-section is also measured as a function of the estimated fraction
of the photon momentum carried by the parton entering the hard
sub-process, {\xg}, with full unfolding for detector effects. Angular
distributions in di-jet events are measured and used to demonstrate
the dominance of quark and gluon initiated processes in different
regions of phase space.

At {\ee} colliders the photons are emitted by the beam
electrons\footnote{Positrons are also referred to as electrons.}.
Most of these photons carry only a small negative four-momentum squared,
$Q^2$, and can be considered quasi-real ($Q^2 \approx 0$). The
electrons are hence scattered with very small polar angles
and are not detected.  Events where one or both scattered electrons
are detected are not considered in the present analysis.  Three
processes contribute to di-jet production in photon-photon collisions:
the direct process where two bare photons interact, the
single-resolved process where a bare photon picks out a parton (quark
or gluon) of the other photon, and the double-resolved process where
partons of both photons interact~\cite{bib-llsmith}.  This separation
is only unambiguous at leading order. At higher orders it becomes
dependent on the process scales.

%
\section{The OPAL detector}
\label{sec-opal}
A detailed description of the OPAL detector can be found
in~\cite{opaltechnicalpaper}. Only the main features relevant to
the present analysis will be given here.
 
The central tracking system is located inside a solenoidal magnet
which provides a uniform axial magnetic field of 0.435~T along the
beam axis.  The magnet is surrounded in the barrel region
($|\cos\theta|<0.82$) by a lead glass electromagnetic calorimeter
(ECAL) and a hadronic sampling calorimeter (HCAL).  Outside the HCAL,
the detector is surrounded by muon chambers. There are similar layers
of detectors in the endcaps ($0.82<|\cos\theta|<0.98$).  The small
angle region from 47 to 140 mrad around the beam pipe on both sides of
the interaction point is covered by the forward calorimeters (FD) and
the region from 33 to 59~mrad by the silicon tungsten luminometers
(SW).

Starting with the innermost components, the tracking system
consists of a high precision silicon micro-vertex detector, a
vertex drift chamber, a large volume jet chamber with 159 layers of
axial anode wires and a set of $z$ chambers measuring the track
coordinates along the beam direction.

The barrel and endcap sections of the ECAL are both constructed
from lead glass blocks with a depth of 24.6 radiation lengths in
the barrel region and more than 22 radiation lengths in the
endcaps. The HCAL consists of streamer tubes and thin multiwire chambers 
instrumenting the gaps in the iron yoke of the magnet, which provides
the absorber material of 4 or more interaction lengths.

  The FD consists of two cylindrical lead-scintillator
calorimeters with a depth of 24 radiation lengths divided
azimuthally into 16 segments.  The SW
detectors consist of 19 layers of silicon detectors and 18 layers
of tungsten, corresponding to a total of 22 radiation lengths.

\section{Monte Carlo simulation}
\label{sec-mc}
The MC generators PYTHIA 5.722~\cite{bib-pythia, bib-schuler} and
PHOJET 1.10~\cite{bib-phojet} are used to study detector effects.
PYTHIA is based on leading order (LO) QCD matrix elements for massless
quarks with added parton showers and hadronisation. PHOJET also
simulates hard interactions through perturbative QCD in LO, but
includes soft interactions through Regge phenomenology before the
partons are hadronised. The probability of finding a parton in the
photon is taken from parametrisations of the parton distribution
functions.  The default choices of {\sas1d}~\cite{bib-sas} for PYTHIA
and LO {\grv}~\cite{bib-grv} for PHOJET are taken for the samples used
to study detector effects.

An increased flow of transverse energy, {\et}, apparently not directly
related to the hard subprocess has been observed in photon-hadron
scattering~\cite{bib-hera}, and has been labeled the underlying event.
Both PHOJET and PYTHIA include a model of multiple parton interactions
(MIA) to simulate such effects. In PYTHIA the amount of MIA added to
the event is controlled by a lower cutoff parameter $\PTMIA$, which
describes the transverse momentum of the parton involved. Following
the studies carried out in~{\cite{bib-opalgg,bib-djopold}}, $\PTMIA$
is set to 1.4~GeV for the {\sas1d} parton densities. In PHOJET the
default setting for MIA is used.

Three non-signal processes are important: hadronic decays of the
$\Zzero$, where initial state photon radiation has reduced the
centre-of-mass energy of the hadronic final state to be close to the 
$\Zzero$ mass, $\gamma\gamma \to \tau\tau$ reactions, and
photon-photon collisions where one of the photons is virtual ({\gsg})
but the scattered electron is not detected. The hadronic $\Zzero$
decays are simulated using PYTHIA. The pair-production of $\tau$-leptons
in photon-photon collisions is simulated using BDK~\cite{bib-bdk}.
Deep-inelastic electron-photon scattering events are studied with the
HERWIG 5.9~\cite{bib-herwig} generator.

All signal and background MC samples were generated with
the full simulation of the OPAL detector~\cite{bib-gopal}. They are
analysed using the same reconstruction algorithms as are applied to
the data.

\section{Definition of di-jet observables}
\label{sec-defobs}

All cross-section measurements use jets reconstructed with the
inclusive {\kp}-clustering algorithm as proposed
in~{\cite{bib-ktclus}} with $R_0=1$. In addition, a cone jet
algorithm~{\cite{bib-coneopal}} with a cone size of 1.0 in
$\eta$-$\phi$-space is employed to study the dependence of the jet
structure on the algorithm used.
A di-jet event is defined as an event with at least two jets
fulfilling the requirements detailed below. In events with more than
two jets, only the two jets with the highest {\etjet} values are
taken.

The primary intentions of this analysis are to study the ability of QCD
theory to describe jet production in photon-photon collisions, and to
explore the photon structure revealed in these hadronic
interactions. The most advanced theoretical predictions to date are
provided by fixed order perturbative calculations up to NLO for the
production of di-jets. These calculations need as input a scale that
is in principle arbitrary, but commonly set to a value related to the
hardness of the interaction. Possible choices for di-jet
production are for example the mean transverse energy {\etmean} or the
maximum {\etjet} of the di-jet system.

The separation of quasi-real and virtual photons is somewhat arbitrary
and therefore needs to be defined. For this analysis we choose values
of $Q^2<4.5$~GeV$^2$ to define quasi-real photons. It is the same
value as used in previous analyses~\cite{bib-opalgg,bib-djopold},
motivated by the acceptance of the low angle calorimeters. The median
$Q^2$ resulting from this definition cannot be determined with the
data since the scattered electrons are not tagged. For the kinematic
range of this analysis both PHOJET and PYTHIA predict the median $Q^2$
to be of the order $10^{-4}$~GeV$^2$.

\subsection{Properties of di-jet events}

In LO QCD, neglecting multiple parton interactions, two hard parton
jets are produced in {\gg} interactions.  In single- or
double-resolved interactions, these jets are expected
to be accompanied by one or two remnant jets.  A pair of variables,
{\xgp} and {\xgm}, can be defined~\cite{bib-LEP2} that estimate
the fraction of the photon's momentum participating in the hard
scattering:
\begin{equation}
\xgp \equiv \frac{\displaystyle{\sum_{\rm jets=1,2}
 (E^\mathrm{jet}+p_z^\mathrm{jet})}}
 {{\displaystyle\sum_{\rm hfs}(E+p_z)}} \;\;\;\mbox{and}\;\;\;
\xgm \equiv \frac{\displaystyle{\sum_{\rm jets=1,2}
 (E^\mathrm{jet}-p_z^\mathrm{jet})}}
{\displaystyle{\sum_{\rm hfs}(E-p_z)}},
\label{eq-xgpm}
\end{equation}
where $p_z$ is the momentum component along the $z$ axis of the
detector and $E$ is the energy of the jets or objects of the hadronic
final state (hfs).  In LO, for direct events, all energy of the event
is contained in two jets, i.e.,~${\xgp}=1$ and ${\xgm}=1$, whereas for
single-resolved or double-resolved events one or both values are
smaller than~1. The di-jet differential cross-section as a function of
{\xg} is therefore particularly well suited to study the structure of
the photon, since it separates predominantly direct events at high
{\xg} ($\xg>0.75$) from predominantly resolved events at low {\xg}
($\xg<0.75$). The fraction of direct, single-resolved and
double-resolved events as a function of $\xg$ predicted by PYTHIA is
shown in Figure~\ref{fig:drdcomp} (a)-(c). The dominance of resolved events
for $\xg<0.75$ is clearly visible.  In these distributions and in the
definitions below, $\xg$ indicates that each event enters the
distribution twice, at the value of {\xgp} and the value of {\xgm}.

Due to the different nature of the underlying partonic process one
expects different distributions of the angle $\Theta^{*}$ between
the jet axis and the axis of the incoming partons or direct photons
in the di-jet centre-of-mass frame. The leading order direct
process $\gamma\gamma \rightarrow \mathrm{q\bar{q}}$ proceeds via
the $t$-channel exchange of a spin-$\frac{1}{2}$ quark, which
leads to an angular dependence $\propto \left( 1- \mathrm{cos}^{2}
\Theta^{*}\right)^{-1}$. In double resolved processes the sum of
all matrix elements, including a large contribution from spin-$1$
gluon exchange, leads to an approximate angular dependence
$\propto \left( 1- |\mathrm{cos}
\Theta^{*}|\right)^{-2}$~\cite{bib-cost}. The contribution of the
different processes to all resolved events depends on the
parton distribution functions of the photon.
An estimator of the angle $\Theta^{*}$ can be formed from the
pseudo-rapidities of the two jets as 
\begin{eqnarray}
\mathrm{cos}\Theta^{*} &=& \mathrm{tanh}\left(
\frac{\eta_{1}^{\mathrm{jet}} - \eta_{2}^{\mathrm{jet}}}{2}\right),
\label{eqn-cost}
\end{eqnarray}
where it is assumed that the jets are collinear in $\phi$ and have
equal transverse energy. Only {\costhst} can be measured, as the
ordering of the jets in the detector is arbitrary. To obtain an
unbiased distribution of {\costhst} the measurement needs to be
restricted to the region where the di-jet invariant mass
$M_{\mathrm{jj}} = 2\etmean/\sqrt{1-\costhst^2}$ is not influenced by
the cuts on {\etjet}~\cite{bib-djopold}. In the present
analysis a cut of $M_{\mathrm{jj}} > 15$~GeV ensures that the {\costhst}
distribution is not biased by the restrictions on {\etjet} for the
range {\costhst}$<$0.8 and ${\etajmean}=|
\left({\eta_{1}^{\mathrm{jet}} + \eta_{2}^{\mathrm{jet}}}\right)/2| <
1$ confines the measurement to the region where the detector
resolution on {\costhst} is good.

\subsection{Differential di-jet cross-sections}

The following differential cross-sections are measured, where the
labels 1 and 2 refer to the two jets with highest {\etjet} in the
event, defined by the {\kp} algorithm:
\begin{eqnarray}
{\dsdetm} & & \mathrm{with}~\,
              \etmean \equiv 
              \frac{E_{\rm T,1}^{\rm jet}+E_{\rm T,2}^{\rm jet}}{2}
              ~~\mathrm{and}~\etmean > 5~\mathrm{GeV}\label{eqn-xsetm}\\
{\dsdxg} & &  \mathrm{in~3~bins~of}~\, 
              \etmean~\left[5-7-11-25\right]
              ~\mathrm{GeV} \label{eqn-xsxg}\\
{\dsdlxg} & & \mathrm{for}~~ 5~\mathrm{GeV} < \etmean < 
              7~\mathrm{GeV}\label{eqn-xslxg}\\
{\dsdetac},\, {\dsdetaf},\,  {\dsddeta} & &  
                          \mathrm{for}~~ \etmean > 5~\mathrm{GeV}
                          \label{eqn-xseta}\\
{\dscost} & & \mathrm{for}~~ \etmean > 5~\mathrm{GeV},~
                           \etajmean < 1,~ M_{\mathrm{jj}} > 15~\mathrm{GeV}
                         \label{eqn-xscost}\\
& & \nonumber\\
& & \mathrm{with~in~all~cases}\nonumber\\
& & \nonumber\\
|\eta^{\mathrm{jet}}_{1,2}|< 2  & &  \mathrm{and~~~~~}
                   \frac{|E_{\rm T,1}^{\rm jet}-E_{\rm T,2}^{\rm 
                   jet}|}
                   {E_{\rm T,1}^{\rm jet}+E_{\rm T,2}^{\rm jet}}
                    < \frac{1}{4} 
                   \label{eqn-asym}
\end{eqnarray}

\noindent
Here, {\etajetc} and {\etajetf} denote the jet with the smaller and larger
value of {\etajet} respectively, and {\detajet} is defined to be the
absolute distance in pseudo-rapidity between the two leading jets.

The combination of the second condition in Equation~({\ref{eqn-asym}})
with the minimum $\etmean$ requirement defines asymmetric $\etjet$
thresholds for the two jets of the di-jet system, which is important
in comparisons to NLO QCD calculations~{\cite{bib-asym}. This method
of defining asymmetric thresholds has previously been used
in~{\cite{bib-h1ppdj}}.

Four regions in {\xgp}-{\xgm}-space are considered (see Figure
\ref{fig:drdcomp} (d)): (A) the complete {\xgp}-{\xgm}-space (full
{\xgpm} range), (B) both {\xgp} and {\xgm} larger than 0.75
(${\xgpm}>0.75$), (C) either {\xgp} or {\xgm} smaller than 0.75
({\xgp} or {\xgm} $< 0.75$), (D) both {\xgp} and {\xgm} smaller than
0.75 (${\xgpm}<0.75$).

The cross-sections (\ref{eqn-xsetm}), (\ref{eqn-xsxg}) and
(\ref{eqn-xslxg}) are measured in regions (A), (C) and (D).  For the
cross-sections in (\ref{eqn-xseta}) regions (C) and (D) are
considered.  The cross-section as a function of {\costhst} in
(\ref{eqn-xscost}) is measured in regions (B) and (D).

\subsection{Jet structure in di-jet events}

The internal structure of jets is studied using the jet shape, which
is defined as the fractional transverse jet energy contained in a
subcone of radius $r$ concentric
with the jet axis, averaged over all jets of the event sample:
\begin{equation}
\psi(r) \equiv \frac{1}{N_{\mathrm{jets}}} \sum_{\mathrm{jets}}
\frac{E\mathrm{_T^{jet}}(r)}{E\mathrm{_T^{jet}}(r=1.0)} 
~~~\mathrm{with}~~~
r=\sqrt{(\Delta\eta)^2+(\Delta\phi)^2}.
\label{shape:def}
\end{equation}
$N_{\mathrm{jet}}$ is the total number of jets analysed. Both {\kp}
and cone jets are analysed in this way. As proposed in
\cite{bib-seym}, only particles assigned to the jet by the jet finders
are considered. Events entering the jet shape distributions are
required to have at least two jets with a transverse energy 3~GeV $<
{\etjet} <$ 20~GeV and a pseudo-rapidity ${\etajet} < 2$. The cone jet
algorithm is used in addition to the {\kp}-clustering algorithm to
demonstrate the different structure of the cone jets with respect to
those defined by the {\kp}-clustering algorithm.

The jet shape is measured in the two regions of {\xgp}-{\xgm}-
space, ${\xgpm} < 0.75$ and ${\xgpm} > 0.75$, in four bins of
{\etjet} with bin boundaries at 3, 6, 9, 12 and 20~GeV
and four bins of {\etajet} between 0 and 2, each bin 0.5 units wide.

\section{Event selection}
\label{sec-evsel}
In this analysis, a sum over all particles in the event or in a jet
means a sum over two kinds of objects: tracks and calorimeter
clusters, including the FD and SW calorimeters.  A track is required
to have a minimum transverse momentum of 120~MeV and at least 20 hits
in the central jet chamber.  The point of closest approach to the
origin must have a distance of less than 25~cm in $z$
and a radial distance of less than 2~cm to the $z$-axis.  Calorimeter
clusters have to pass an energy threshold of 100~MeV in the barrel
section or 250~MeV in the endcap section for the ECAL, 600~MeV for the
barrel and endcap sections of the HCAL, 1~GeV for the FD, and 2~GeV for
the SW.  An algorithm is applied to avoid double-counting of particle
momenta in the central tracking system and their energy deposits in
the calorimeters~\cite{bib-opalgg}. The measured hadronic final state
for each event consists of all objects thus defined.

Di-jet events are preselected using the {\kp} algorithm by requiring
at least two jets with ${\etajet}<2$ and a transverse energy
${\etjet}>3$~GeV.  Photon-photon scattering events are selected using
the requirements detailed below.  The corresponding distributions in
Figure~\ref{fig:selq1} compare the sum of the simulated signal and
background processes to the data, uncorrected for detector effects.
For each distribution shown, all selection criteria are applied except
the one on the quantity plotted.  The signal MC generators PHOJET and
PYTHIA are found to underestimate the cross-section by about 20\% in
these comparisons, and are scaled up accordingly.  Of all non-signal
processes studied, only those listed in Section \ref{sec-mc}
contribute significantly. Comparisons of the rate of di-jet events in
photon-photon collisions where one of the photons is virtual (see for
example ~\cite{bib-rooke}) show that the prediction of the MC
generator used is too low by about a factor of two. The prediction of
the contribution from {\gsg} events has been scaled up accordingly.

All distributions are sufficiently well described by the sum of signal
and background contributions.  The total contribution of non-signal
processes to the selected event sample is about 5\% after the
following selection criteria have been applied:

\begin{itemize}
  
\item The sum of all energy deposits in the ECAL and HCAL
  (Figure~\ref{fig:selq1}\,(a)) has to be less than 55~GeV to remove
  background from hadronic {\Zzero}-decays in events with a radiative
  return to the {\Zzero}-peak.
  
\item The visible invariant mass measured in the ECAL,
  $W_{\mathrm{ECAL}}$, has to be greater than 3~GeV to suppress low
  energy events. 
  
\item The missing transverse momentum of the event, {\ptmiss},
  calculated from the measured hadronic final state, has to be less
  than \mbox{$0.05\cdot{\ebeam}$}.
  
\item At least 7 tracks must have been found in the tracking chambers.
  This cut reduces mostly the contamination from $\gamma\gamma \to
  \tau\tau$ events.  The distribution of the track multiplicity is
  shown in Figure~\ref{fig:selq1}\,(b). The discrepancy in shape
  between data and simulation is not present when using PYTHIA
  instead of PHOJET as signal MC generator, and is addressed in the
  study of model dependencies in Section~\ref{sec-syst}.
  
\item To remove events with scattered electrons in the FD or in the SW
  calorimeters, the total sum of the energy measured in the FD has to
  be less than \mbox{$0.25\cdot{\ebeam}$} and the total sum of the
  energy measured in the SW calorimeter has to be less than
  \mbox{$0.18\cdot{\ebeam}$}. These cuts also reduce the contamination
  from hadronic {\Zzero}-decays with their thrust axis close to the
  beam direction. The energy sum in the FD calorimeter scaled by the
  beam energy is shown in Figure~\ref{fig:selq1}\,(c).
  
\item The $z$ position of the primary vertex is required to 
  satisfy $|z|<5$~cm and the net charge $Q$ of the event calculated
  from adding the charges of all tracks is required to be $|Q| \le
  5$ to reduce background due to beam-gas interactions. 

\item To remove events originating from interactions between beam
  electrons and the beam-pipe the radial distance of the primary
  vertex from the beam axis has to be less than 3~cm.
  
\item To further reject background from hadronic {\Zzero}-decays and
  from deep-inelas\-tic electron-photon scattering an invariant mass,
  {\mj1h2}, is calculated from the jet with highest {\etjet} in the
  event and the four-vector sum of all hadronic final state objects in
  the hemisphere opposite to the direction defined by this jet. The
  quantity {\mj1h2} is a simple reconstruction of the {\Zzero}-mass in
  case of background from hadronic {\Zzero}-decays, and will therefore
  be larger on average for this type of background than for signal
  events. Events with {\mj1h2} $>$ 55~GeV are rejected. The
  distribution of {\mj1h2} is shown in Figure~\ref{fig:selq1}\,(d).

\end{itemize}

We use data at centre-of-mass energies {\sqee} from 189~GeV to
209~GeV.  For the purpose of this analysis, the difference between the
data taken at the various values of {\sqee} is small and therefore the
distributions for all energies have been added.  The luminosity
weighted average centre-of-mass energy {\sqee} is approximately
198.5~GeV.  The efficiency to trigger di-jet events in the region of
phase space considered in this analysis has been shown to be close to
100\%~\cite{bib-djopold}.  

\section{Transverse energy flow in di-jet events}
\label{sec-etflow}
NLO QCD calculations do not take into account the possibility of
an underlying event which leads to an increased {\et}-flow and
therefore to an increased jet cross-section above a given threshold of
{\etjet}.  In PYTHIA and PHOJET the underlying event is simulated by
multiple parton interactions.  The contribution from multiple parton
interactions is not known a priori, but has to be adjusted to give a
good description of the data. In this analysis the size of this
contribution is taken from our previous study of di-jet events
in~\cite{bib-djopold}. The transverse energy flow from an underlying
event is expected to be small compared to the transverse energy of the
leading jets, and it is not correlated to the direction of the jet
axes.  The energy flow outside the jets will therefore be most
sensitive to the presence of an underlying event~\cite{bib-h1dijet}.
Additional energy outside the leading jets will shift the {\xg}
distributions towards lower values.

To study the performance of the MC generators in describing the energy
flow several uncorrected distributions are used.  The average
{\et}-flow per event is measured with respect to the jet axis as a
function of $\Delta\phi$ and {$\Delta\etah$}. The variable {\etah} is
equivalent to {$\eta$}, except that it is signed positively if {\xgp}
is greater than or equal {\xgm}, and signed negatively otherwise. The
definition of {\etah} ensures that the energy flow associated with the
``more resolved'' photon, i.e., the smaller value of {\xg}, will
always appear on the left hand side of the plots.  The profiles in
$\Delta\phi$ consider a range of $|\Delta\eta| = 1$ around the
jet-axis, while a $|\Delta\phi|$-range of $\pi/2$ around the jet-axis
is considered for the profiles in {$\Delta\etah$}.  The two leading
jets in {\etjet} in each event are considered. Another sensitive
variable is the energy flow with the leading two jets in the event
removed, $E^{out}$, as a function of {$\etah$} .  All objects are
excluded inside a cone of radius 1.3 in ${\eta}$-${\phi}$ around the
two leading jets.

Multiple parton interactions are important in interactions where the
photon is resolved. It is therefore interesting to study separately
the three cases of (a) two resolved photons, (b) one resolved photon,
or (c) no resolved photon in the interaction.  Experimentally these
situations can be approximated by choosing events with ${\xgpm} <
0.75$, ${\xgp}$ or ${\xgm} < 0.75$, or ${\xgpm} > 0.75$,
i.e. regions (D), (C) and (B) defined in Section~\ref{sec-defobs}.

Figures~{\ref{fig:jprophieta}} and {\ref{fig:etout}} show the
jet profiles and the energy flow outside the leading jets.  The data
are compared to a mixture of signal (PHOJET or PYTHIA) and background
MC simulation.  The contributions of signal and background are
weighted according to their cross-section in each region of phase
space. The background MC generators used are the same as in
Figure~\ref{fig:selq1}.

For the {$\phi$}-profiles in Figure~{\ref{fig:jprophieta}} it is
evident that both PHOJET and PYTHIA are able to describe the data in
the region of the high {\et} jets around zero. Moving away from the
jet axis PHOJET predicts an energy flow which is too low compared
to the data, especially for ${\xgpm}<0.75$. This corresponds to the
area where effects from an underlying event are expected to be most
prominent.  PHOJET improves towards higher {\xg}.  PYTHIA reproduces
the data reasonably well. Such differences between PHOJET and PYTHIA
are not evident in the {\etah}-profiles.  Here in the case of resolved
photons the {\et}-flow is dominated by the photon remnant(s), and is
reasonably well described by both generators. The jets entering
Figure~{\ref{fig:jprophieta}} are selected from the range $10 <
{\etjet} < 25$~GeV. No significant deviation from the behaviour just
described is observed when selecting ${\etjet}<10$~GeV. 

The energy-flow outside the two
leading jets is shown in Figure~{\ref{fig:etout}}. Again PYTHIA
describes the data well, while PHOJET is somewhat low.  With both
models used to unfold the data as a systematic check, we conclude that
the details of the energy flow around the two leading jets are
sufficiently well under control and remaining influences are included
in the systematic uncertainty of the cross-section and jet shape
measurements.

\section{Data corrections}
\label{sec-cross}

An example of the uncorrected distributions as a function of {\xg}
including the contribution of the remaining background events is shown
in Figure~{\ref{fig:obsdet}}.  To obtain jet cross-sections which can
be compared to theoretical calculations, we use MC simulations to
correct for the selection cuts, the resolution effects of the detector
and the background from non-signal processes. Backgrounds are first
subtracted bin-by-bin from all distributions. For the differential
cross-sections as a function of {\etmean} and {\xg}, sizable migration
and resolution effects are to be expected. We therefore apply a matrix
unfolding method, as implemented in the GURU program~\cite{bib-guru},
for these distributions.  The results are cross checked using a
bin-by-bin correction.  By definition {\xg} can only take values
between zero and unity. At either extremity no adjacent bins are
available. To avoid instabilities expected from the smoothing
procedures in the unfolding for the first and last bin, the central
values for these bins are taken from the bin-by-bin correction method.
The {\costhst}, {\etajet} and {\detajet} distributions are corrected
bin-by-bin, as only small migrations are expected here. PHOJET is used
as the default signal MC generator for the unfolding.

The correction method employed for the jet shapes is a bin-by-bin
correction using the MC simulations to correct for detector
effects.  The contribution of the same background processes as for
the cross-section measurements was studied. The influence on the
signal was found to be less than 1\%. Therefore the subtraction of
the background was omitted in this analysis. Both PYTHIA and PHOJET
were used to estimate the correction factors to study their model
dependence.

\section{Systematic uncertainties}
\label{sec-syst}

The overall systematic uncertainty is determined from the sources
listed below added in quadrature. The same sources are considered
for the measurement of the differential cross-sections and the
jet shapes, with the exception of the background, which has been
neglected for the jet shapes as discussed in
Section~\ref{sec-cross}.

\begin{itemize}
  
\item To assess the uncertainty associated with the subtraction of
  background events, the predictions for hadronic decays of the
  $\Zzero$ and for $\gamma\gamma \to \tau\tau$ reactions are
  conservatively varied by $10\%$ without contributing
  significantly to the systematic error. The prediction of the
  contribution from {\gsg} events has been scaled up by a factor of
  two as described above.  By comparing the predictions to the data
  for large $E_\mathrm{FD}/E_\mathrm{BEAM}$ and $M_\mathrm{J1H2}$ (see
  Figure~\ref{fig:selq1}), where this background dominates, we
  determine that this scaling factor can be varied by no more than
  about 30\% in order to keep a good description of the data.  The
  scaling factor is varied accordingly. The uncertainty from all
  the background subtraction is typically 2-4\%.
   
\item To estimate the systematic error arising from the specific
  model used for the unfolding, both PYTHIA and PHOJET are used to
  unfold the data. The estimated uncertainty derived from this
  study is typically 10\%, and up to 20\% in some cases for the
  differential cross-sections, and 1-2\% for the jet shapes.
  
\item The absolute energy scale of the ECAL
  calorimeter is known to about 3\%~\cite{bib-totxs}
  for the range of jet energies in this analysis. To estimate the
  influence on the observables the energy scale is varied by this
  amount and the analysis is repeated. The cross-sections change by
  5-10\% due to this variation. The estimated uncertainty for the
  jet shapes is about 1\%.
  
\item The selection criteria described in Section~\ref{sec-evsel} are
  varied simultaneously both to be more restrictive and to allow more
  events into the analysis to exclude a strong dependence on the event
  selection. Selection criteria based on energy measurements are
  varied by 10\% of their central value, which is considered
  conservative given the uncertainty in the energy scale and the
  energy resolution of the calorimeter. The number of tracks required
  and the maximum net charge of the event are changed by $\pm1$. The
  allowed radial distance and $z$ position of the primary vertex
  are varied by 0.5~cm and 1~cm respectively.  The uncertainty on the
  cross-section derived from all these variations is typically 5-10\%,
  and up to 20\% in some cases for the differential cross-section, and
  about 2-4\% for the jet shapes.

\end{itemize} 

The uncertainty on the determination of the integrated luminosity is
much less than 1\%, and is neglected. For the differential
cross-sections the systematic uncertainties evaluated for each bin were
averaged with the results from its two neighbours (single neighbour
for endpoints) to reduce the effect of bin-to-bin fluctuations.

\section{Hadronisation corrections}
\label{sec-hadr}

The differential di-jet cross-sections measured are compared to NLO
QCD calculations which predict jet cross-sections for partons, whereas
the experimental jet cross-sections are presented for hadrons.
Effects due to the modelling of the hadronisation process are not
taken into account in the NLO calculation.  Because the partons in the
MC generators and the partons in the NLO calculations are defined in
different ways there is as yet no rigorous procedure to use the MC
generators to correct the data so that they can be compared to the NLO
parton level predictions. However, as the MC generators are the only
available option so far, they are used to study the approximate size
of these hadronisation corrections. For this purpose the prediction of
the MC generators at the level of the partonic final state is
calculated and divided by the prediction obtained from the hadronic
final state. The resulting correction factor is labeled {\hadcor}. The
partonic final state consists of all partons at the end of the parton
shower. The hadronic final state utilises all charged and neutral
particles with lifetimes greater than $3\times 10^{-10}$~s, which are
treated as stable.

Examples of hadronisation corrections estimated by PYTHIA 6.161 and
HERWIG 6.1 for the observables defined in Section~\ref{sec-defobs} are
shown in Figure~{\ref{fig:hc}}. The numerical values can be found
in~\cite{bib-hepdata}. In PYTHIA the partonic final state is
hadronised according to the string fragmentation model, while HERWIG
uses cluster fragmentation. In all plots the full {\xgp}-{\xgm}-range
is considered. The theoretical calculations are corrected bin-by-bin
using the mean of the hadronisation corrections estimated using PYTHIA
and HERWIG.

Figure~{\ref{fig:hc}}~(a) shows {\hadcor} as a function of {\etmean}.
The correction is less than 10\% for {\etmean} greater than about
10~GeV, but rises to about 25\% for PYTHIA and 15\% for HERWIG towards
small {\etmean}. The corrections for observbles involving the jet
pseudo-rapidities are dominated by the low {\etmean} region. They are
essentially flat and around 20\% for PYTHIA. HERWIG estimates these
corrections to be around 10\%.

Figure~{\ref{fig:hc}}~(b) shows {\hadcor} as a function of {\xg} for
the lowest bin in {\etmean} defined in Section~\ref{sec-defobs}. From
the figure it is evident that hadronisation causes large corrections
for ${\xg}>0.75$. The effect is reduced for higher values of
{\etmean}, as can be seen in Figure~{\ref{fig:hc}}~(c) and (d). The
large corrections for ${\xg}>0.75$ are mainly due to the large
influence hadronisation has on the distribution of direct events,
which are peaked at ${\xg}=1$ for the partonic final state of the LO
calculation, but are much more smeared out at the level of stable
hadrons (see Figure~{\ref{fig:drdcomp}} (a)-(c)). While both the
measurement and the NLO calculation are perfectly valid for the
presented bin sizes, the hadronisation corrections needed for the
comparison introduces a large migration between the two bins above
${\xg}=0.75$. For a sensible comparison in this region one should
therefore consider the sum of the two bins above ${\xg}=0.75$ and
compare it to the corresponding sum for the data.

\section{Results}
\label{sec-results}

\subsection{Jet structure in di-jet events}

In Figure~\ref{fig:jsh01}\,(a) the jet shape, $\Psi(r)$, is shown for
the {\kp} algorithm for both ${\xgpm} > 0.75$ and ${\xgpm} < 0.75$.
The first sample is dominated by direct photon-photon interactions and
hence by quark-initiated jets. As is demonstrated in the figure, jets
in this sample are more collimated than for small values of {\xgpm},
where the cross-section is dominated by resolved processes and hence
has a large contribution from gluon-initiated jets. In both cases the
jets become more collimated with increasing transverse energy, as is
shown in Figure~\ref{fig:jsh01}\,(c). There is no significant
dependence on the jet pseudo-rapidity (Figure~\ref{fig:jsh01}\,(d)).
Both PHOJET and PYTHIA give an adequate description of the jet shapes
as can be seen in Figures~\ref{fig:jsh01}\,(b), (c), and (d).

Figure~\ref{fig:jsh02} compares the shapes of jets defined by the cone
algorithm and the {\kp} algorithm, in each case compared to the
shape as obtained from PYTHIA. As for the {\kp}-jets, the
jets defined by the cone algorithm are more collimated in the
quark-dominated sample and always become more collimated for
increasing transverse energy, while there is no dependence on the jet
pseudo-rapidity. The cone-jets are significantly broader than the jets
defined by the {\kp} algorithm at low {\etjet}. With increasing
{\etjet}, jets become more collimated and the two jet algorithms give
similar results.  While the {\kp}-jets are well described by PYTHIA
and PHOJET, the jet shapes obtained for the cone-jets are somewhat
broader than in the data.

\subsection{Differential di-jet cross-sections}

Only the {\kp} jet algorithm is used for the measurement of the
differential di-jet cross-sections. The experimental results are
compared to a perturbative QCD calculation at NLO~\cite{bib-ggnlo}
which uses the {\grvnlo} parametrisation of the parton
distribution functions of the photon~\cite{bib-grv}, and was
repeated for the kinematic conditions of the present analysis. The
renormalisation and factorisation scales are set to the maximum
$\etjet$ in the event.  The calculation was performed in the
$\overline{\mathrm{MS}}$-scheme with five light flavours and
$\Lambda^{(5)}_{\mathrm{QCD}}=130$~MeV. The average of the
hadronisation corrections estimated by PYTHIA and HERWIG have
been applied to the calculation for this comparison.

Figure~{\ref{fig:costhst}} and Table~{\ref{tab:costhst}} show the
differential di-jet cross-section as a function of {\costhst} for both
${\xgpm} > 0.75$ and ${\xgpm} < 0.75$. The steeper rise with
increasing {\costhst} from the dominating spin-$1$ gluon exchange in
the second sample is clearly visible (see Section~\ref{sec-defobs}).
The shape of both samples is well described by NLO QCD. For
${\xgpm} < 0.75$ the NLO calculation is about 20\,\% below the data. It
should be noted that in this region the contribution from the
underlying event, not included in the calculation, is expected to be
largest, as discussed in more detail below. For ${\xgpm} > 0.75$ the
NLO QCD prediction is about 20\,\% above the data. While here the
contribution from MIA is small, this region is affected by rather
large hadronisation corrections as discussed in
Section~\ref{sec-hadr}, which translates into an uncertainty of the
normalisation in comparing the theoretical prediction to the data.

The differential di-jet cross-section as a function of the mean
transverse energy $\etmean$ of the di-jet system is shown in
Figure~{\ref{fig:etmxs}} and Table~{\ref{tab:etmxs}}.  At high
{\etmean} the cross-section is expected to be dominated by direct
processes, associated with the region ${\xgpm} > 0.75$.  Consequently
we observe a significantly softer spectrum for the case ${\xgpm} <
0.75$ than for the full {\xgp}-{\xgm}-space. The calculation is in
good agreement with the data for the full {\xgp}-{\xgm}-range and for
{\xgp} or {\xgm} $< 0.75$. The cross-section predicted for ${\xgpm} <
0.75$ is again below the measurement. PYTHIA 6.161 is in good
agreement with the measured distributions using the {\sas1d} parton
densities.

Figure~{\ref{fig:xgxs}} and Tables~{\ref{tab:xgxs}}
and~{\ref{tab:xgxslg}} show the di-jet cross-section as a function of
{\xg} and {\lxg}.  The cross-section for the lowest values of
{\etmean} shows the largest fraction of events at $\xg<0.75$ of the
three ranges considered, and is hence most sensitive to
gluon-initiated processes. The di-jet cross-section logarithmic in
{\xg} emphasises the region of lowest accessible {\xg}, which extends
down to approximately 0.02.

As $\etmean$ increases, the fraction of events with ${\xg} > 0.75$
increases. In MC simulations these are predominantly direct events.
The sensitivity to the gluon density in the photon is hence expected
to decrease with increasing {\etmean}.  On the other hand, NLO QCD
predictions which use {\etmean} as the process-relevant scale are
expected to become more reliable as this scale increases.  It is hence
important to provide measurements at both low and high values of
{\etmean}, to study all aspects of the theory.

PYTHIA using {\sas1d} is in good agreement with the measured
distributions, with a tendency to be too low for small values of {\xg}.
The shaded histogram at the bottom of each plot indicates the MIA
contribution to the PYTHIA prediction. The numerical values of this
contibution can be found in~\cite{bib-hepdata}. NLO QCD predicts the
shape of the cross-sections well for $\xg<0.75$, but is too low by
about 10-20\,\% especially at low {\etmean}. As MIA are not included
in this calculation it is interesting to note that the MIA
contribution to the cross-section as obtained from PYTHIA is
similar in size to the discrepancy.

The region of ${\xg} > 0.75$ suffers from large hadronisation
corrections as discussed in Section~\ref{sec-hadr}. The uncertainty
for the data-theory comparison associated with these large corrections
can be reduced by considering the sum of the two bins above ${\xg} =
0.75$, for which NLO QCD indeed gives an adequate description
of the data.

Figure~{\ref{fig:nlocont}} demonstrates the effect of using different
parton distribution functions of the photon on the NLO QCD prediction.
{\afgnlo}~\cite{bib-afg} and {\gs96nlo}~\cite{bib-gs96} are used in
addition to the default {\grvnlo}.  The sensitivity of the
cross-section to the different gluon density in each case is clearly
visible for the gg-contribution (Figure~{\ref{fig:nlocont}} (b)), but
is less pronounced for the full cross-section as can be seen in
Figure~{\ref{fig:nlocont}} (a), due to compensating effects from
processes involving the quark densities (Figures~{\ref{fig:nlocont}}
(c) and (d)). A global analysis, beyond the scope of this paper, using
for example $F_2^\gamma$ measurements to constrain simultaneously the
quark densities hence promises to yield the highest sensitivity to the
gluon density in the photon.

In Figure~{\ref{fig:xgxssr}} (Tables~{\ref{tab:xgxssr}}
and~{\ref{tab:xgxslg}}) the same cross-sections as in
Figure~{\ref{fig:xgxs}} are shown for the case {\xgp} or {\xgm} $<
0.75$.  Here the cross-section is dominated by interactions where one
of the two incident photons is resolved. The multiple parton
interactions used in PYTHIA to model an underlying event are much
suppressed in this case, and NLO QCD describes both the shape and
normalisation of the data well.  The opposite effect can be observed
in Figure~{\ref{fig:xgxsdr}} (Tables~{\ref{tab:xgxsdr}}
and~{\ref{tab:xgxslg}}), where for the case ${\xgpm} < 0.75$ one
expects a large influence of multiple parton interactions, as
demonstrated again by the shaded histogram at the bottom of each plot.
The cross-sections change by as much as 50\% in the low {\etmean}
region, when MIA are switched on. For higher {\etmean} the influence
is not as strong. Even with MIA switched on, PYTHIA using {\sas1d} is
too low. The deficit visible in the normalisation of the NLO
calculation is again of similar size as the MIA contribution to the
cross-section obtained from PYTHIA.

Complementary information can be obtained by measuring the angular
distributions of the two highest {\etjet} jets in di-jet events.  The
{\etajet} and {\detajet} dependence of the di-jet cross-section is
dominated by the low {\etjet} events.  The cross-sections measured are
listed in Tables~{\ref{tab:d-etaxs}},~{\ref{tab:c-etaxs}}
and~{\ref{tab:f-etaxs}}. In Figure~{\ref{fig:etaxssr}} the di-jet
cross-sections as a function of {\detajet}, {\etajetc} and {\etajetf}
are shown for the case {\xgp} or {\xgm} $< 0.75$. Again the multiple
parton interactions used in PYTHIA to model an underlying event are
much suppressed in this case.  PYTHIA using {\sas1d} is about 20\% too
low. The prediction of NLO QCD is in good agreement with the data in
both shape and normalisation.

In Figure~{\ref{fig:etaxsdr}} the same cross-sections are presented
for the case of ${\xgpm}<0.75$.  As expected, the effect of including
MIA in PYTHIA is again sizable.  When multiple parton interactions are
switched on, the prediction obtained from PYTHIA reproduces the data
reasonably well. Again the prediction of NLO QCD is too
low by about the size of the MIA contribution to the cross-section
obtained by PYTHIA.

\section{Conclusions}
\label{sec-sum}
We have studied di-jet production in photon-photon interactions with
the OPAL detector at {\ee} centre-of-mass energies {\sqee} from 189 to
209~GeV with an integrated luminosity of 593~{\ipb}.  The data are
combined into one sample with a luminosity weighted average
centre-of-mass energy of approximately $\sqee = 198.5$~GeV.
Jets are reconstructed using an inclusive {\kp}-clustering algorithm
for the measurement of differential di-jet cross-sections, and using
both the inclusive {\kp} and a cone algorithm for the study of jet
structure.

Jet shapes, $\Psi(r)$, have been studied in two separate samples:
${\xgpm}>0.75$, which is dominated by direct processes and hence by
quark-initiated jets, and ${\xgpm}<0.75$, dominated by resolved events
and therefore by gluon-initiated jets. As expected from QCD the jets
in the first sample are significantly more collimated than for
${\xgpm}<0.75$. Jets in both samples become more collimated with
increasing transverse energy, but show no significant dependence on
the jet pseudo-rapidity. Jets defined by the cone algorithm are
substantially broader than those defined by the {\kp} algorithm at low
{\etjet}. However the difference  decreases with increasing {\etjet}.
The shape of {\kp}-jets is well described by PYTHIA and PHOJET. The
jet shapes obtained for the cone-jets are somewhat broader in
PYTHIA and PHOJET than in the data.

Inclusive differential di-jet cross-sections have been measured as a
function of {\costhst}, {\etmean}, {\etajet} and {\detajet} and, for
the first time, as a function of {\xg} in several bins of {\etmean}.
Different regions of the {\xgp}-{\xgm}-space are explored to separate
experimentally direct from resolved interactions and to study and
control the influence of an underlying event.  By measuring the
cross-sections for events in which either {\xgp} or {\xgm} is smaller
than 0.75 we have isolated a region of phase space in which resolved
photon processes dominate, and which at the same time is much less
sensitive to multiple parton interactions. By performing the
measurement also for ${\xgpm}<0.75$, observables are made available
which are sensitive to the amount of multiple parton interactions
added in the prediction, and which can be used to study these effects
in detail.

A strong rise with increasing {\costhst} is observed for the
differential di-jet cross-section for ${\xgpm}<0.75$, as expected from
QCD for a sample with a significant contribution from spin-$1$ gluon
exchange. The flatter distribution for direct events is also in good
agreement with the QCD calculation.

The differential di-jet cross-sections as a function of {\etmean},
{\etajet}, {\detajet} and {\xg} are in good agreement with the
next-to-leading order perturbative QCD calculation except for
${\xgpm}<0.75$, where the calculation is too low. As this calculation
does not include a model for the underlying event that is expected to be
largest in this region, it is interesting to note that the discrepancy
is of similar size to the contribution of multiple parton interactions
to the PYTHIA prediction. The sensitivity of the results presented to
the gluon density in the photon is clearly visible in NLO QCD
predictions using different parton distribution functions, but is
compensated to some extent by anti-correlated differences in the
respective quark-densities. A global analysis using additional data
sets to simultaneously constrain the quark densities hence promises to
yield the highest sensitivity to the gluon density in the photon.

The measurements carried out for events in which only either {\xgp} or
{\xgm} is smaller than 0.75 are a unique data set. While this region
is almost insensitive to multiple parton interactions, the fraction of
events at small {\xg} is still sizable, which indicates a significant
contribution of resolved processes and hence a good sensitivity to the
hadronic structure of the photon.  The good agreement of data and
theory in this region in particular confirms that perturbative QCD in
next-to-leading order is able to describe correctly the inclusive
production of di-jets in photon-photon collisions.

%
%
\appendix
\par
\section*{Acknowledgements}
\par
We thank M.{\thinspace}Klasen and collaborators for providing the NLO
QCD calculations and D.{\thinspace}Berge and S.{\thinspace}K\"onig for
their contribution to the analysis of jet shapes presented in this
paper.  

We particularly wish to thank the SL Division for the
efficient operation of the LEP accelerator at all energies and for
their close cooperation with our experimental group.  In addition to
the support staff at our own
institutions we are pleased to acknowledge the  \\
Department of Energy, USA, \\
National Science Foundation, USA, \\
Particle Physics and Astronomy Research Council, UK, \\
Natural Sciences and Engineering Research Council, Canada, \\
Israel Science Foundation, administered by the Israel
Academy of Science and Humanities, \\
Benoziyo Center for High Energy Physics,\\
Japanese Ministry of Education, Culture, Sports, Science and
Technology (MEXT) and a grant under the MEXT International
Science Research Program,\\
Japanese Society for the Promotion of Science (JSPS),\\
German Israeli Bi-national Science Foundation (GIF), \\
Bundesministerium f\"ur Bildung und Forschung, Germany, \\
National Research Council of Canada, \\
Hungarian Foundation for Scientific Research, OTKA T-029328,
and T-038240,\\
The NWO/NATO Fund for Scientific Reasearch, the Netherlands.\\

%
%

\cleardoublepage
\bibliography{djpr}

\begin{thebibliography}{10}

\bibitem{bib-amy}
{AMY Collaboration, B.J.{\thinspace}Kim et~al.}{,}
\newblock Phys.~Lett.~B325 (1994) 248.

\bibitem{bib-topaz}
{TOPAZ Collaboration, H.{\thinspace}Hayashii et~al.}{,}
\newblock Proceedings of Photon '95, Sheffield, UK, 8-13 April 1995, edited by
  D.J.{\thinspace}Miller, S.L.{\thinspace}Cartwright and V.{\thinspace}Khoze,
  World Scientific (Singapore) 1995, p133;\\ TOPAZ Collaboration,
  H.{\thinspace}Hayashii et~al., Phys.~Lett.~B314 (1993) 149.

\bibitem{bib-opalgg}
{OPAL Collaboration, K.{\thinspace}Ackerstaff et~al.}{,}
\newblock Z.~Phys.~C73 (1997) 433.

\bibitem{bib-djopold}
{OPAL Collaboration, G.{\thinspace}Abbiendi et~al.}{,}
\newblock Eur.~Phys.~J.~C10 (1999) 547.

\bibitem{bib-ktclus}
{S.{\thinspace}Catani, Yu.L.{\thinspace}Dokshitzer, M.H.{\thinspace}Seymour and
  B.R.{\thinspace}Webber}{,}
\newblock Nucl.~Phys.~B406 (1993) 187;\\ S.D.{\thinspace}Ellis,
  D.E.{\thinspace}Soper, Phys.~Rev.~D48 (1993) 3160.

\bibitem{bib-coneopal}
{OPAL Collaboration, R.{\thinspace}Akers et~al.}{,}
\newblock Z.~Phys.~C63 (1994) 197.

\bibitem{bib-ktisbest}
{M.{\thinspace}Wobisch and T.{\thinspace}Wengler}{,}
\newblock hep-ph/9907280;\\ M.H.{\thinspace}Seymour, hep-ph/9707349;\\
  S.D.{\thinspace}Ellis, Z.{\thinspace}Kunszt and D.E.{\thinspace}Soper,
  Phys.~Rev.~Lett. 69 (1992) 3615.

\bibitem{bib-llsmith}
{C.H.{\thinspace}Llewellyn{\thinspace}Smith}{,}
\newblock Phys.~Lett.~B79 (1978) 83.

\bibitem{opaltechnicalpaper}
{OPAL Collaboration, K.{\thinspace}Ahmet et~al.}{,}
\newblock Nucl.~Instrum.~Methods~A305 (1991) 275;\\ S.{\thinspace}Anderson
  et~al., Nucl.~Instrum.~Methods~A403 (1998) 326;\\ OPAL Collaboration,
  G.{\thinspace}Abbiendi et~al., Eur.~Phys.~J. C14 (2000) 373.

\bibitem{bib-pythia}
{T.{\thinspace}Sj\"ostrand}{,}
\newblock Comp.~Phys.~Comm.~82 (1994) 74;\\ T.{\thinspace}Sj\"ostrand, LUND
  University Report, LU-TP-95-20 (1995).

\bibitem{bib-schuler}
{G.A.{\thinspace}Schuler and T.{\thinspace}Sj\"ostrand}{,}
\newblock Z.~Phys.~C73 (1997) 677;\\ G.A.{\thinspace}Schuler and
  T.{\thinspace}Sj\"ostrand, Nucl.~Phys.~B407 (1993) 539.

\bibitem{bib-phojet}
{R.{\thinspace}Engel}{,}
\newblock Z.~Phys.~C66 (1995) 203;\\ R.{\thinspace}Engel and
  J.{\thinspace}Ranft, Phys.~Rev.~D54 (1996) 4244.

\bibitem{bib-sas}
{G.A.{\thinspace}Schuler and T.{\thinspace}Sj\"ostrand}{,}
\newblock Z.~Phys.~C68 (1995) 607.

\bibitem{bib-grv}
{M.{\thinspace}Gl\"uck, E.{\thinspace}Reya and A.{\thinspace}Vogt}{,}
\newblock Phys.~Rev.~D45 (1992) 3986;\\ M.{\thinspace}Gl\"uck,
  E.{\thinspace}Reya and A.{\thinspace}Vogt, Phys.~Rev.~D46 (1992) 1973.

\bibitem{bib-hera}
{H1 Collaboration, C.{\thinspace}Adloff et~al.}{,}
\newblock Eur.~Phys.~J.~C1 (1998) 97;\\ ZEUS Collaboration,
  J.{\thinspace}Breitweg et~al., Eur.~Phys.~J.~C1 (1998) 109;\\ ZEUS
  Collaboration, J.{\thinspace}Breitweg et~al., Eur.~Phys.~J.~C4 (1998) 591;\\
  ZEUS Collaboration, J.{\thinspace}Breitweg et~al., Eur.~Phys.~J.~C11 (1999)
  35.

\bibitem{bib-bdk}
{F.A.{\thinspace}Berends, P.H.{\thinspace}Daverveldt and
  R.{\thinspace}Kleiss}{,}
\newblock Nucl.~Phys.~B253 (1985) 421;\\ F.A.{\thinspace}Berends,
  P.H.{\thinspace}Daverveldt and R.{\thinspace}Kleiss, Comp.~Phys.~Comm.~40
  (1986) 271, 285 and 309.

\bibitem{bib-herwig}
{G.{\thinspace}Marchesini et~al.}{,}
\newblock Comp.~Phys.~Comm.~67 (1992) 465;\\ G.{\thinspace}Corcella et~al.,
  JHEP 0101 (2001) 010.

\bibitem{bib-gopal}
{J.{\thinspace}Allison et~al.}{,}
\newblock Nucl.~Instrum.~Methods A317 (1992) 47.

\bibitem{bib-LEP2}
{L.{\thinspace}L\"onnblad and M.{\thinspace}Seymour (convenors)}{,}
\newblock {\em $\gamma\gamma$ Event Generators}, in ``Physics at LEP2'', CERN
  96-01, eds.~G.{\thinspace}Altarelli, T.{\thinspace}Sj\"ostrand and
  F.{\thinspace}Zwirner, Vol.~2 (1996) 187.

\bibitem{bib-cost}
{for example: H.{\thinspace}Kolanoski}{,}
\newblock {\em Two-Photon Physics at e$^+$e$^-$ Storage Rings}, Springer-Verlag
  (1984);\\ B.L.{\thinspace}Combridge, J.{\thinspace}Kripfganz and
  J.{\thinspace}Ranft, Phys.~Lett.~B70 (1977) 234;\\ D.W.{\thinspace}Duke and
  J.F.{\thinspace}Owens, Phys.~Rev.~D26 (1982) 1600.

\bibitem{bib-asym}
{M.{\thinspace}Klasen and G.{\thinspace}Kramer}{,}
\newblock Phys.~Lett.~B366 (1996) 385;\\ S.{\thinspace}Frixione and
  G.{\thinspace}Ridolfi, Nucl.~Phys.~B507 (1997) 315.

\bibitem{bib-h1ppdj}
{H1 Collaboration, C. Adloff et al.}{,}
\newblock Eur. Phys. J.~C1 (1998) 97.

\bibitem{bib-seym}
{M.H.{\thinspace}Seymour}{,}
\newblock Nucl.~Phys.~B513 (1998) 269.

\bibitem{bib-rooke}
{A.M.{\thinspace}Rooke (for the OPAL collaboration)}{,}
\newblock Proceedings of Photon '97, Egmond aan Zee, The Netherlands, 10-18 May
  1997, edited by A.Buijs and F.C.Ern\'e, World Scientific (Singapore) 1997,
  p465;\\ A.M.{\thinspace}Rooke, University of London, PhD Thesis, September
  1998.

\bibitem{bib-h1dijet}
{H1 Collaboration, S.{\thinspace}Aid et~al.}{,}
\newblock Z.~Phys.~C70 (1996) 17.

\bibitem{bib-guru}
{A.{\thinspace}H\"{o}cker, V.{\thinspace}Kartvelishvili}{,}
\newblock Nucl.~Instrum.~Methods~A372 (1996) 469.

\bibitem{bib-totxs}
{OPAL Collaboration, G.{\thinspace}Abbiendi et~al.}{,}
\newblock Eur.~Phys.~J.~C14 (2000) 199.

\bibitem{bib-hepdata}
{HEPDATA: The Durham RAL Databases}{,}
\newblock Durham Database Group, at Durham University(UK),
  http://www-spires.dur.ac.uk/HEPDATA.

\bibitem{bib-ggnlo}
{M.{\thinspace}Klasen, T.{\thinspace}Kleinwort and G.{\thinspace}Kramer}{,}
\newblock Eur.~Phys.~J.~Direct~C1 (1998) 1;\\ B.{\thinspace}P\"otter,
  Eur.~Phys.~J.~Direct~C5 (1999) 1.

\bibitem{bib-afg}
{P.{\thinspace}Aurenche, J.P.{\thinspace}Guillet and
  M.{\thinspace}Fontannaz}{,}
\newblock Z.~Phys.~C64 (1994) 621.

\bibitem{bib-gs96}
{L.E.{\thinspace}Gordon and J.K.{\thinspace}Storrow}{,}
\newblock Nucl.~Phys.~B489 (1997) 405.

\end{thebibliography}
\cleardoublepage
%
%
%
\renewcommand{\arraystretch}{1.20}
\begin{table}[t]
\begin{center}
\begin{tabular}{|ccc|c|c|c|}
\hline
\multicolumn{3}{|c|}{ {\costhst}} &
{$\dscost$} ~[pb]&{$\delta_\mathrm{\,stat}$ ~[pb]}&
{$\delta_\mathrm{\,sys}$} ~[pb]\\
%
\hline
\hline
\multicolumn{6}{|c|}{${\xgpm} > 0.75$ }\\
\hline
%
   0.0 &  -- &  0.1 &   5.04 &   0.49 &   0.32 \\ 
   0.1 &  -- &  0.2 &   4.96 &   0.49 &   0.34 \\ 
   0.2 &  -- &  0.3 &   5.22 &   0.49 &   0.32 \\ 
   0.3 &  -- &  0.4 &   5.35 &   0.48 &   0.38 \\ 
   0.4 &  -- &  0.5 &   4.90 &   0.45 &   0.32 \\ 
   0.5 &  -- &  0.6 &   6.73 &   0.56 &   0.47 \\ 
   0.6 &  -- &  0.7 &   7.09 &   0.54 &   0.43 \\ 
   0.7 &  -- &  0.8 &   8.42 &   0.61 &   0.51 \\ 
%
\hline
\hline
\multicolumn{6}{|c|}{${\xgpm} < 0.75$ }\\
\hline
%
   0.0 &  -- &  0.1 &   1.50 &   0.31 &   0.24 \\ 
   0.1 &  -- &  0.2 &   2.63 &   0.46 &   0.36 \\ 
   0.2 &  -- &  0.3 &   2.91 &   0.50 &   0.41 \\ 
   0.3 &  -- &  0.4 &   3.73 &   0.62 &   0.43 \\ 
   0.4 &  -- &  0.5 &   4.36 &   0.61 &   0.62 \\ 
   0.5 &  -- &  0.6 &   6.52 &   0.80 &   0.74 \\ 
   0.6 &  -- &  0.7 &   9.90 &   0.92 &   1.07 \\ 
   0.7 &  -- &  0.8 &  13.05 &   0.96 &   1.02 \\ 
%
\hline

\end{tabular}
\end{center}
\caption{The di-jet cross-section as a function of {\costhst} for 
  the two regions in
  {\xgp}-{\xgm}-space indicated in the table. The total uncertainty
  for each bin is the quadratic sum of the statistical
  and systematic uncertainty given in the table.}
\label{tab:costhst}
\end{table}
%
%
\renewcommand{\arraystretch}{1.20}
\begin{table}[t]
\begin{center}
\begin{tabular}{|ccc|c|c|c|}
\hline
\multicolumn{3}{|c|}{ {\etmean} ~[GeV]} &
{$\dsdetm$} ~[pb/GeV]&{$\delta_\mathrm{\,stat}$ ~[pb/GeV]}&
{$\delta_\mathrm{\,sys}$} ~[pb/GeV]\\
%
\hline
\hline
\multicolumn{6}{|c|}{full~ {\xgp} - {\xgm} -- range }\\
\hline
%
   5.00 &  -- &  6.54 &  11.54 &   0.36 &   0.59 \\ 
   6.54 &  -- &  8.55 &   4.16 &   0.16 &   0.25 \\ 
   8.55 &  -- & 11.18 &   1.45 &   0.07 &   0.09 \\ 
  11.18 &  -- & 14.62 &  0.543 &  0.036 &  0.044 \\ 
  14.62 &  -- & 19.12 &  0.176 &  0.020 &  0.020 \\ 
  19.12 &  -- & 25.00 & 0.0607 & 0.0117 & 0.0087 \\ 
%
\hline
\hline
\multicolumn{6}{|c|}{${\xgp}$ ~or~ 
                     ${\xgm} < 0.75$ }\\
\hline
%
   5.00 &  -- &  6.54 &   3.56 &   0.21 &   0.22 \\ 
   6.54 &  -- &  8.55 &   1.32 &   0.09 &   0.10 \\ 
   8.55 &  -- & 11.18 &  0.475 &  0.045 &  0.037 \\ 
  11.18 &  -- & 14.62 &  0.190 &  0.025 &  0.024 \\ 
  14.62 &  -- & 19.12 & 0.0518 & 0.0127 & 0.0073 \\ 
%
\hline
\hline
\multicolumn{6}{|c|}{${\xgpm} < 0.75$ }\\
\hline
%
   5.00 &  -- &  6.54 &   5.94 &   0.41 &   0.66 \\ 
   6.54 &  -- &  8.55 &   1.84 &   0.16 &   0.23 \\ 
   8.55 &  -- & 11.18 &  0.461 &  0.060 &  0.071 \\ 
  11.18 &  -- & 14.62 & 0.0781 & 0.0192 & 0.0130 \\ 
%
\hline
\end{tabular}
\end{center}
\caption{The di-jet cross-section as a function of the mean 
  transverse energy $\etmean$ of the di-jet system, for the three
  regions in {\xgp}-{\xgm}-space indicated in the table.}
\label{tab:etmxs}
\end{table}
%
%
\renewcommand{\arraystretch}{1.20}
\begin{table}[t]
\begin{center}
\begin{tabular}{|ccc|c|c|c|}
\hline
\multicolumn{3}{|c|}{ {\xg}} &
{$\dsdxg$} ~[pb]&{$\delta_\mathrm{\,stat}$ ~[pb]}&
{$\delta_\mathrm{\,sys}$} ~[pb]\\
%
\hline
\hline
\multicolumn{6}{|c|}{full~ {\xgp} - {\xgm} -- range }\\
\hline
\hline
\multicolumn{6}{|c|}{ 5 GeV $<$ {\etmean} $<$ 7 GeV}\\
\hline
%
   0.000 &  -- &  0.125 &  47.29 &   2.51 &   4.41 \\ 
   0.125 &  -- &  0.250 &  46.29 &   2.73 &   4.69 \\ 
   0.250 &  -- &  0.375 &  33.66 &   2.16 &   3.37 \\ 
   0.375 &  -- &  0.500 &  25.47 &   1.93 &   2.25 \\ 
   0.500 &  -- &  0.625 &  25.70 &   1.86 &   2.16 \\ 
   0.625 &  -- &  0.750 &  38.91 &   2.47 &   2.70 \\ 
   0.750 &  -- &  0.875 &  79.59 &   4.07 &   8.37 \\ 
   0.875 &  -- &  1.000 &  51.97 &   5.12 &   5.49 \\ 
 %
\hline
\hline
\multicolumn{6}{|c|}{ 7 GeV $<$ {\etmean} $<$ 11 GeV}\\
\hline
%
   0.000 &  -- &  0.125 &  12.41 &   1.37 &   1.70 \\ 
   0.125 &  -- &  0.250 &  16.78 &   1.55 &   2.05 \\ 
   0.250 &  -- &  0.375 &  12.64 &   1.34 &   1.47 \\ 
   0.375 &  -- &  0.500 &  10.36 &   1.16 &   0.86 \\ 
   0.500 &  -- &  0.625 &  11.67 &   1.24 &   0.95 \\ 
   0.625 &  -- &  0.750 &  14.80 &   1.53 &   1.05 \\ 
   0.750 &  -- &  0.875 &  38.84 &   2.72 &   3.99 \\ 
   0.875 &  -- &  1.000 &  43.07 &   3.00 &   4.75 \\ 
%
\hline
\hline
\multicolumn{6}{|c|}{ 11 GeV $<$ {\etmean} $<$ 25 GeV}\\
\hline
%
   0.000 &  -- &  0.125 &   1.51 &   0.53 &   0.39 \\ 
   0.125 &  -- &  0.250 &   2.83 &   0.52 &   0.64 \\ 
   0.250 &  -- &  0.375 &   2.78 &   0.54 &   0.43 \\ 
   0.375 &  -- &  0.500 &   2.30 &   0.48 &   0.36 \\ 
   0.500 &  -- &  0.625 &   2.97 &   0.57 &   0.41 \\ 
   0.625 &  -- &  0.750 &   4.43 &   0.66 &   0.55 \\ 
   0.750 &  -- &  0.875 &  11.04 &   1.23 &   0.94 \\ 
   0.875 &  -- &  1.000 &  26.81 &   1.85 &   2.09 \\ 
%
\hline
\end{tabular}
\end{center}

\caption{The di-jet cross-section as a function of $\xg$ for the 
regions of the mean transverse energy $\etmean$ of the di-jet system 
and the region in {\xgp}-{\xgm}-space indicated in the table.}
\label{tab:xgxs}
\end{table}
%
%
\renewcommand{\arraystretch}{1.20}
\begin{table}[t]
\begin{center}
\begin{tabular}{|ccc|c|c|c|}
\hline
\multicolumn{3}{|c|}{ {\xg}} &
{$\dsdxg$} ~[pb]&{$\delta_\mathrm{\,stat}$ ~[pb]}&
{$\delta_\mathrm{\,sys}$} ~[pb]\\
%
\hline
\multicolumn{6}{|c|}{${\xgp}$ ~or~ 
                     ${\xgm} < 0.75$  }\\
\hline
\hline
\multicolumn{6}{|c|}{ 5 GeV $<$ {\etmean} $<$ 7 GeV}\\
\hline
%
   0.000 &  -- &  0.125 &  12.47 &   0.96 &   1.18 \\ 
   0.125 &  -- &  0.250 &   8.87 &   0.74 &   1.16 \\ 
   0.250 &  -- &  0.375 &   7.00 &   0.69 &   0.83 \\ 
   0.375 &  -- &  0.500 &   5.12 &   0.58 &   0.65 \\ 
   0.500 &  -- &  0.625 &   5.99 &   0.69 &   0.56 \\ 
   0.625 &  -- &  0.750 &  12.36 &   1.27 &   0.98 \\ 
   0.750 &  -- &  0.875 &  38.50 &   1.92 &   4.19 \\ 
   0.875 &  -- &  1.000 &  11.80 &   1.72 &   1.30 \\ 
%
\hline
\hline
\multicolumn{6}{|c|}{ 7 GeV $<$ {\etmean} $<$ 11 GeV}\\
\hline
%
   0.000 &  -- &  0.125 &   5.44 &   0.73 &   0.62 \\ 
   0.125 &  -- &  0.250 &   4.84 &   0.52 &   0.48 \\ 
   0.250 &  -- &  0.375 &   3.45 &   0.44 &   0.31 \\ 
   0.375 &  -- &  0.500 &   2.91 &   0.41 &   0.21 \\ 
   0.500 &  -- &  0.625 &   3.28 &   0.48 &   0.25 \\ 
   0.625 &  -- &  0.750 &   5.38 &   0.76 &   0.39 \\ 
   0.750 &  -- &  0.875 &  15.03 &   1.17 &   1.94 \\ 
   0.875 &  -- &  1.000 &   8.63 &   1.19 &   1.31 \\ 
\hline
\end{tabular}
\end{center}
\caption{The di-jet cross-section as a function of $\xg$ for the
regions of the mean transverse energy $\etmean$ of the di-jet
system and the region in {\xgp}-{\xgm}-space indicated in the
table. }
\label{tab:xgxssr}
\end{table}
%
%
\renewcommand{\arraystretch}{1.20}
\begin{table}[t]
\begin{center}
\begin{tabular}{|ccc|c|c|c|}
\hline
\multicolumn{3}{|c|}{ {\xg}} &
{$\dsdxg$} ~[pb]&{$\delta_\mathrm{\,stat}$ ~[pb]}&
{$\delta_\mathrm{\,sys}$} ~[pb]\\
%
\hline
\multicolumn{6}{|c|}{${\xgpm} < 0.75$ }\\
\hline
\hline
\multicolumn{6}{|c|}{ 5 GeV $<$ {\etmean} $<$ 7 GeV}\\
\hline
%
   0.000 &  -- &  0.125 &  42.37 &   2.94 &   4.71 \\ 
   0.125 &  -- &  0.250 &  37.66 &   2.44 &   4.32 \\ 
   0.250 &  -- &  0.375 &  26.87 &   1.99 &   2.66 \\ 
   0.375 &  -- &  0.500 &  19.63 &   1.88 &   1.49 \\ 
   0.500 &  -- &  0.625 &  17.84 &   1.55 &   1.10 \\ 
   0.625 &  -- &  0.750 &  26.71 &   3.86 &   1.73 \\ 
%
\hline
\hline
\multicolumn{6}{|c|}{ 7 GeV $<$ {\etmean} $<$ 11 GeV}\\
\hline
%
   0.000 &  -- &  0.125 &   8.81 &   1.08 &   1.54 \\ 
   0.125 &  -- &  0.250 &  11.00 &   1.00 &   1.72 \\ 
   0.250 &  -- &  0.375 &  10.30 &   0.96 &   1.24 \\ 
   0.375 &  -- &  0.500 &   8.10 &   0.81 &   0.76 \\ 
   0.500 &  -- &  0.625 &   7.49 &   0.94 &   0.64 \\ 
   0.625 &  -- &  0.750 &   8.51 &   1.58 &   0.72 \\ 
\hline
\end{tabular}
\end{center}
\caption{The di-jet cross-section as a function of $\xg$ for the
regions of the mean transverse energy $\etmean$ of the di-jet
system and the region in {\xgp}-{\xgm}-space indicated in the
table. }
\label{tab:xgxsdr}
\end{table}
%
%
\renewcommand{\arraystretch}{1.20}
\begin{table}[t]
\begin{center}
\begin{tabular}{|ccc|c|c|c|}
\hline
\multicolumn{3}{|c|}{ {\lxg}} &
{$\dsdlxg$} ~[pb]&{$\delta_\mathrm{\,stat}$ ~[pb]}&
{$\delta_\mathrm{\,sys}$} ~[pb]\\
%
\hline
\hline
\multicolumn{6}{|c|}{ 5 GeV $<$ {\etmean} $<$ 7 GeV}\\
\hline
\hline
\multicolumn{6}{|c|}{full~ {\xgp} - {\xgm} -- range }\\
\hline
%
  -1.65 &  -- & -1.32 &   4.00 &   0.42 &   0.42 \\ 
  -1.32 &  -- & -0.99 &   9.54 &   0.76 &   1.07 \\ 
  -0.99 &  -- & -0.66 &  16.55 &   1.12 &   1.83 \\ 
  -0.66 &  -- & -0.33 &  27.51 &   1.39 &   2.35 \\ 
  -0.33 &  -- &  0.00 &  76.87 &   1.69 &   5.05 \\ 
%
\hline
\hline
\multicolumn{6}{|c|}{${\xgp}$ ~or~ 
                     ${\xgm} < 0.75$ }\\
\hline
%
  -1.65 &  -- & -1.32 &   1.29 &   0.18 &   0.18 \\ 
  -1.32 &  -- & -0.99 &   2.38 &   0.24 &   0.32 \\ 
  -0.99 &  -- & -0.66 &   3.64 &   0.30 &   0.44 \\ 
  -0.66 &  -- & -0.33 &   4.87 &   0.38 &   0.47 \\ 
  -0.33 &  -- &  0.00 &  25.87 &   0.99 &   2.16 \\ 
%
\hline
\hline
\multicolumn{6}{|c|}{${\xgpm} < 0.75$ }\\
\hline
%
  -1.65 &  -- & -1.32 &   3.08 &   0.44 &   0.41 \\ 
  -1.32 &  -- & -0.99 &   7.81 &   0.63 &   1.02 \\ 
  -0.99 &  -- & -0.66 &  13.95 &   0.92 &   1.55 \\ 
  -0.66 &  -- & -0.33 &  19.39 &   1.13 &   1.78 \\ 
  -0.33 &  -- &  0.00 &  18.01 &   1.53 &   1.36 \\ 
%
\hline
\end{tabular}
\end{center}
\caption{The di-jet cross-section as a function of $\lxg$ for the
region of the mean transverse energy $\etmean$ of the di-jet
system and the regions in {\xgp}-{\xgm}-space indicated in the
table. }
\label{tab:xgxslg}
\end{table}
%
%
\renewcommand{\arraystretch}{1.20}
\begin{table}[t]
\begin{center}
\begin{tabular}{|ccc|c|c|c|}
\hline
\multicolumn{3}{|c|}{ {\detajet}} &
{$\dsddeta$} ~[pb]&{$\delta_\mathrm{\,stat}$ ~[pb]}&
{$\delta_\mathrm{\,sys}$} ~[pb]\\
%
\hline
\hline
\multicolumn{6}{|c|}{${\xgp}$ ~or~ 
                     ${\xgm} < 0.75$ }\\
\hline
%
   0.0 &  -- &  0.5 &   6.68 &   0.27 &   0.38 \\ 
   0.5 &  -- &  1.0 &   5.51 &   0.24 &   0.31 \\ 
   1.0 &  -- &  1.5 &   3.98 &   0.20 &   0.23 \\ 
   1.5 &  -- &  2.0 &   2.34 &   0.15 &   0.15 \\ 
   2.0 &  -- &  2.5 &   1.22 &   0.11 &   0.11 \\ 
   2.5 &  -- &  3.0 &  0.631 &  0.081 &  0.074 \\ 
   3.0 &  -- &  3.5 &  0.251 &  0.056 &  0.035 \\ 
%
\hline
\hline
\multicolumn{6}{|c|}{${\xgpm} < 0.75$ }\\
\hline
%
   0.0 &  -- &  0.5 &   7.47 &   0.36 &   0.66 \\ 
   0.5 &  -- &  1.0 &   6.89 &   0.35 &   0.60 \\ 
   1.0 &  -- &  1.5 &   5.42 &   0.29 &   0.46 \\ 
   1.5 &  -- &  2.0 &   3.93 &   0.24 &   0.30 \\ 
   2.0 &  -- &  2.5 &   2.44 &   0.19 &   0.30 \\ 
   2.5 &  -- &  3.0 &   1.43 &   0.16 &   0.24 \\ 
   3.0 &  -- &  3.5 &  0.618 &  0.124 &  0.133 \\ 
%
\hline

\end{tabular}
\end{center}
\caption{The di-jet cross-section as a function of {\detajet} for 
  the two leading jets in {\etjet}, for the two regions in
  {\xgp}-{\xgm}-space indicated in the table.}
\label{tab:d-etaxs}
\end{table}
%
%
\renewcommand{\arraystretch}{1.20}
\begin{table}[t]
\begin{center}
\begin{tabular}{|ccc|c|c|c|}
\hline
\multicolumn{3}{|c|}{ {\etajetc}} &
{$\dsdetac$} ~[pb]&{$\delta_\mathrm{\,stat}$ ~[pb]}&
{$\delta_\mathrm{\,sys}$} ~[pb]\\
%
\hline
\hline
\multicolumn{6}{|c|}{${\xgp}$ ~or~ 
                     ${\xgm} < 0.75$ }\\
\hline
%
   0.00 &  -- &  0.25 &  11.48 &   0.50 &   0.65 \\ 
   0.25 &  -- &  0.50 &   9.25 &   0.44 &   0.55 \\ 
   0.50 &  -- &  0.75 &   7.19 &   0.38 &   0.46 \\ 
   0.75 &  -- &  1.00 &   5.30 &   0.32 &   0.31 \\ 
   1.00 &  -- &  1.25 &   3.49 &   0.26 &   0.23 \\ 
   1.25 &  -- &  1.50 &   2.60 &   0.24 &   0.25 \\ 
   1.50 &  -- &  1.75 &   1.49 &   0.21 &   0.17 \\ 
%
\hline
\hline
\multicolumn{6}{|c|}{${\xgpm} < 0.75$ }\\
\hline
%
   0.00 &  -- &  0.25 &  14.93 &   0.67 &   1.24 \\ 
   0.25 &  -- &  0.50 &  13.38 &   0.65 &   1.17 \\ 
   0.50 &  -- &  0.75 &  10.83 &   0.60 &   0.96 \\ 
   0.75 &  -- &  1.00 &   7.17 &   0.47 &   0.60 \\ 
   1.00 &  -- &  1.25 &   5.59 &   0.46 &   0.54 \\ 
   1.25 &  -- &  1.50 &   3.20 &   0.41 &   0.41 \\ 
   1.50 &  -- &  1.75 &   2.01 &   0.41 &   0.32 \\ 
%
\hline

\end{tabular}
\end{center}
\caption{The di-jet cross-section as a function of {\etajetc} for 
  the two regions in
  {\xgp}-{\xgm}-space indicated in the table.}
\label{tab:c-etaxs}
\end{table}
%
%
\renewcommand{\arraystretch}{1.20}
\begin{table}[t]
\begin{center}
\begin{tabular}{|ccc|c|c|c|}
\hline
\multicolumn{3}{|c|}{ {\etajetf}} &
{$\dsdetaf$} [pb]&{$\delta_\mathrm{stat}$ [pb]}&
{$\delta_\mathrm{sys}$} [pb]\\
%
\hline
\hline
\multicolumn{6}{|c|}{${\xgp}$ ~or~ 
                     ${\xgm} < 0.75$  }\\
\hline
%
   0.00 &  -- &  0.25 &  0.906 &  0.137 &  0.065 \\ 
   0.25 &  -- &  0.50 &   2.68 &   0.23 &   0.18 \\ 
   0.50 &  -- &  0.75 &   4.30 &   0.29 &   0.26 \\ 
   0.75 &  -- &  1.00 &   5.65 &   0.33 &   0.33 \\ 
   1.00 &  -- &  1.25 &   6.18 &   0.34 &   0.33 \\ 
   1.25 &  -- &  1.50 &   6.86 &   0.36 &   0.34 \\ 
   1.50 &  -- &  1.75 &   7.27 &   0.40 &   0.48 \\ 
   1.75 &  -- &  2.00 &   7.03 &   0.44 &   0.55 \\ 
%
\hline
\hline
\multicolumn{6}{|c|}{${\xgpm} < 0.75$ }\\
\hline
%
   0.00 &  -- &  0.25 &   1.03 &   0.17 &   0.16 \\ 
   0.25 &  -- &  0.50 &   4.11 &   0.38 &   0.54 \\ 
   0.50 &  -- &  0.75 &   5.64 &   0.40 &   0.60 \\ 
   0.75 &  -- &  1.00 &   6.71 &   0.41 &   0.51 \\ 
   1.00 &  -- &  1.25 &   8.40 &   0.47 &   0.74 \\ 
   1.25 &  -- &  1.50 &  10.72 &   0.62 &   0.94 \\ 
   1.50 &  -- &  1.75 &  11.30 &   0.68 &   1.18 \\ 
   1.75 &  -- &  2.00 &   9.86 &   0.72 &   0.96 \\ 
%
\hline

\end{tabular}
\end{center}
\caption{The di-jet cross-section as a function of {\etajetf} for 
  the two regions in
  {\xgp}-{\xgm}-space indicated in the table.}
\label{tab:f-etaxs}
\end{table}

\cleardoublepage

\begin{figure}[ht]
\begin{center}
\includegraphics[width=0.95\textwidth]{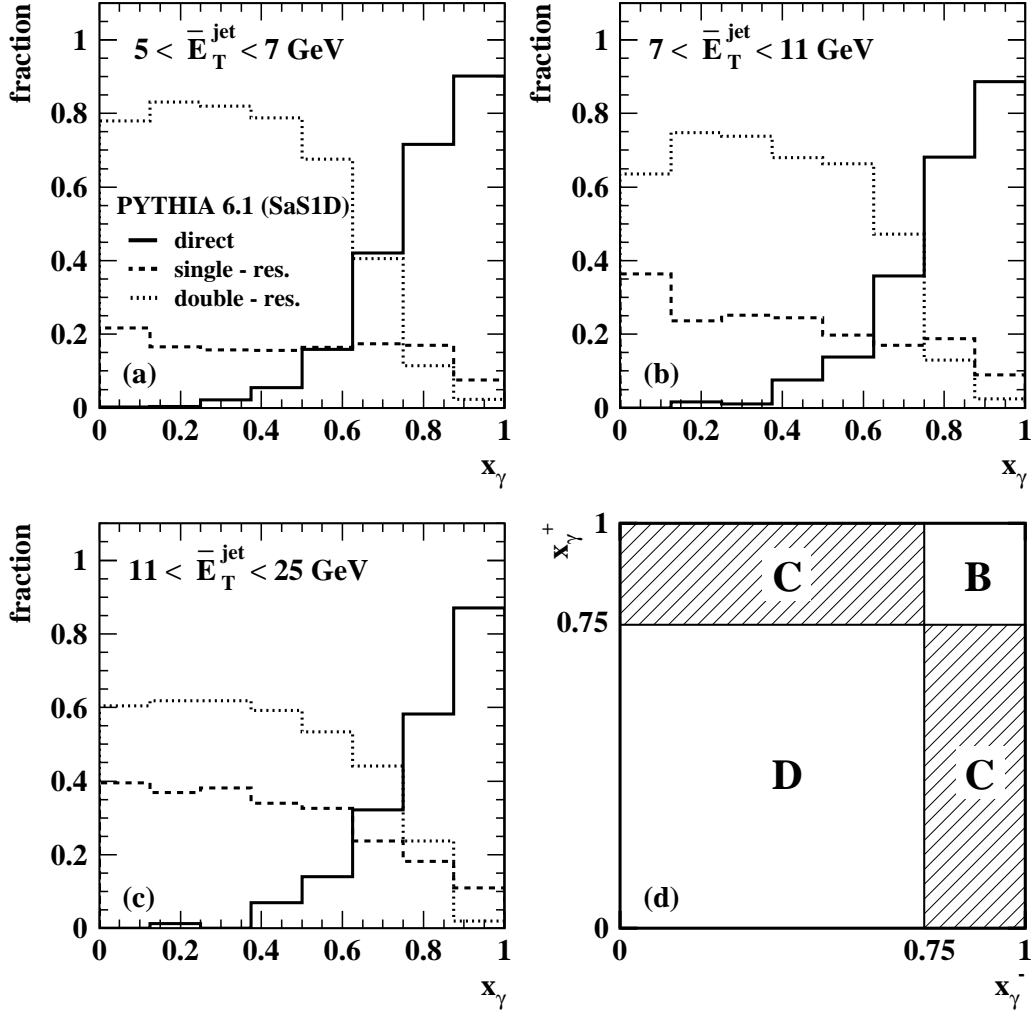}
\end{center}
\caption{ (a)-(c): The relative contribution of direct, 
  single-resolved, and double-resolved processes according to PYTHIA
  at the hadron level for the cross-sections as a function of {\xg}
  for the full {\xgp}-{\xgm}-range (see Section~\ref{sec-defobs}). In
  (d) the regions in {\xgp}-{\xgm}-space used in addition to the full
  {\xgp}-{\xgm}-range (referred to as (A)) are illustrated: (B) both
  {\xgp} and {\xgm} larger than 0.75 (${\xgpm}>0.75$), (C) either
  {\xgp} or {\xgm} smaller than 0.75 ({\xgp} or {\xgm} $< 0.75$), (D)
  both {\xgp} and {\xgm} smaller than 0.75 (${\xgpm}<0.75$).}
\label{fig:drdcomp}
\end{figure}


\begin{figure}[ht]
\begin{center}
\includegraphics[width=0.95\textwidth]{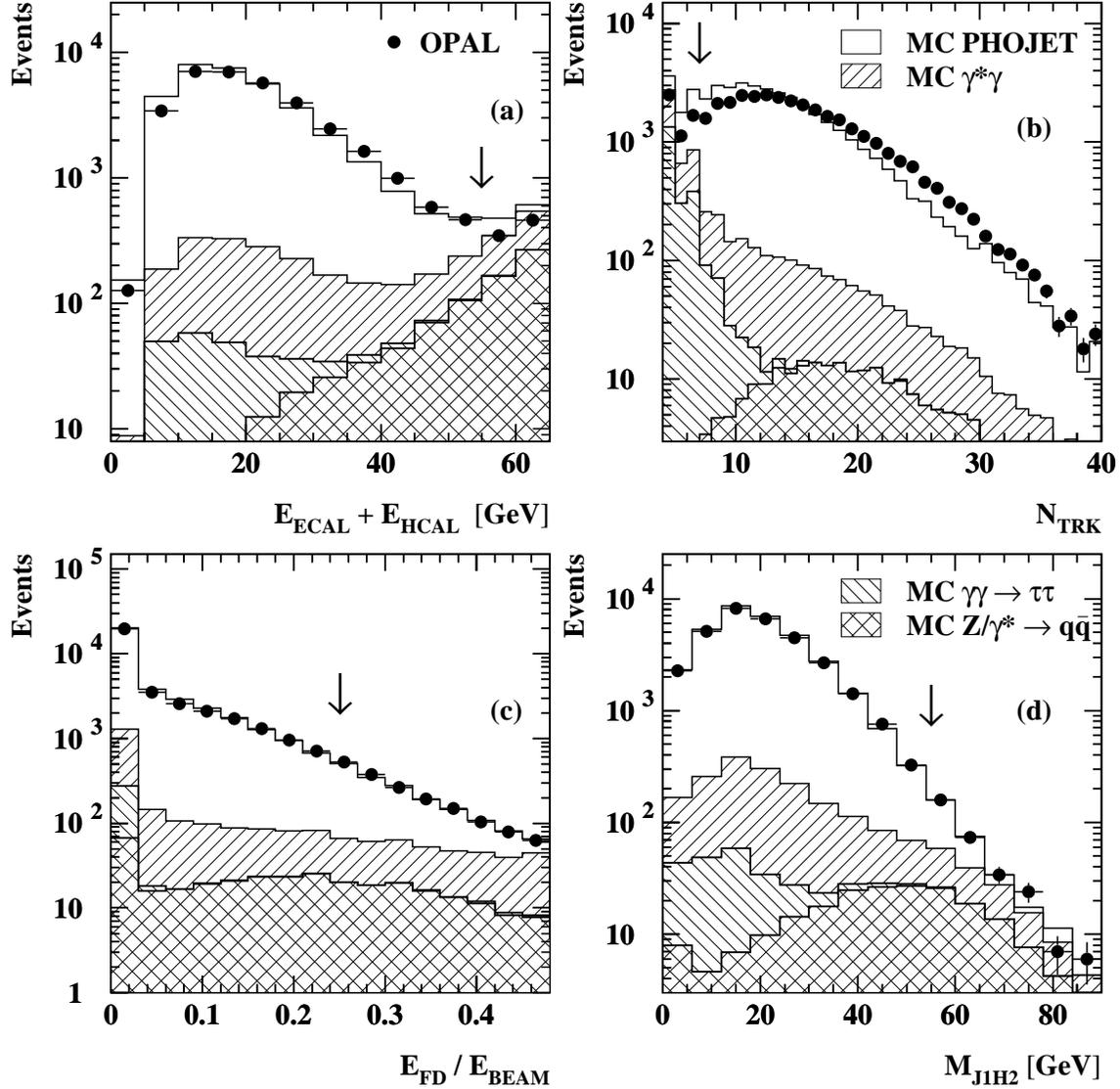}
\end{center}
\caption{ Comparison of event quantities for uncorrected data  with
the simulation for the di-jet sample including contributions from
background processes. (a) shows the sum of energy measured in the
electromagnetic and hadronic calorimeter, (b)  the number of tracks
in the event, (c) is the sum of energy in the FD detector scaled by
the beam energy, and (d) the invariant mass of the jet with the
highest {\etjet} in the event and the four-vector calculated from
all objects in the opposite hemisphere as seen from this jet. The
statistical error is shown where larger than the marker size.  The
label {\gsg} stands for simulated photon-photon collision events in
which one of the photons has a virtuality larger than 4.5~GeV$\,^2$
as discussed in Section~\ref{sec-defobs}.  }
\label{fig:selq1}
\end{figure}


\begin{figure}[ht]
\begin{center}
\includegraphics[width=0.94\textwidth]{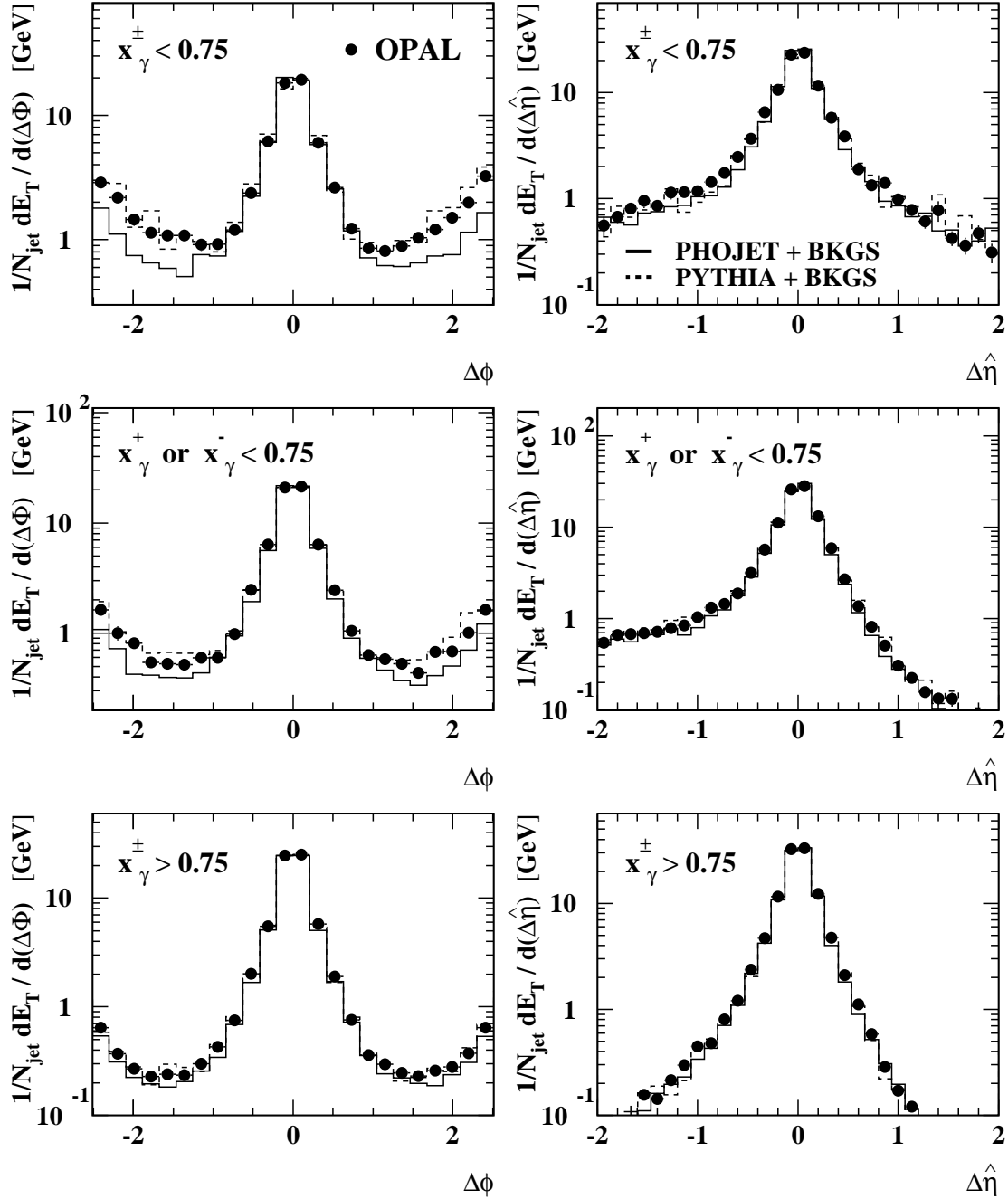}
\end{center}
\caption{Jet profiles: The $\et$-flow normalised to the number of
  jets $N_{jet}$ as a function of the distance from the jet axis in
  $\phi$ and {\etah}. Jets are selected from the range $10 < {\etjet}
  < 25$~GeV. The statistical error is shown where larger than the
  marker size.  The data are compared to a mixture of signal (PHOJET
  or PYTHIA) and background MC simulation.  The background MC
  simulations used are the same as in Figure~\ref{fig:selq1}.}
\label{fig:jprophieta}
\end{figure}

\cleardoublepage


\begin{figure}[ht]
\begin{center}
\includegraphics[width=0.95\textwidth]{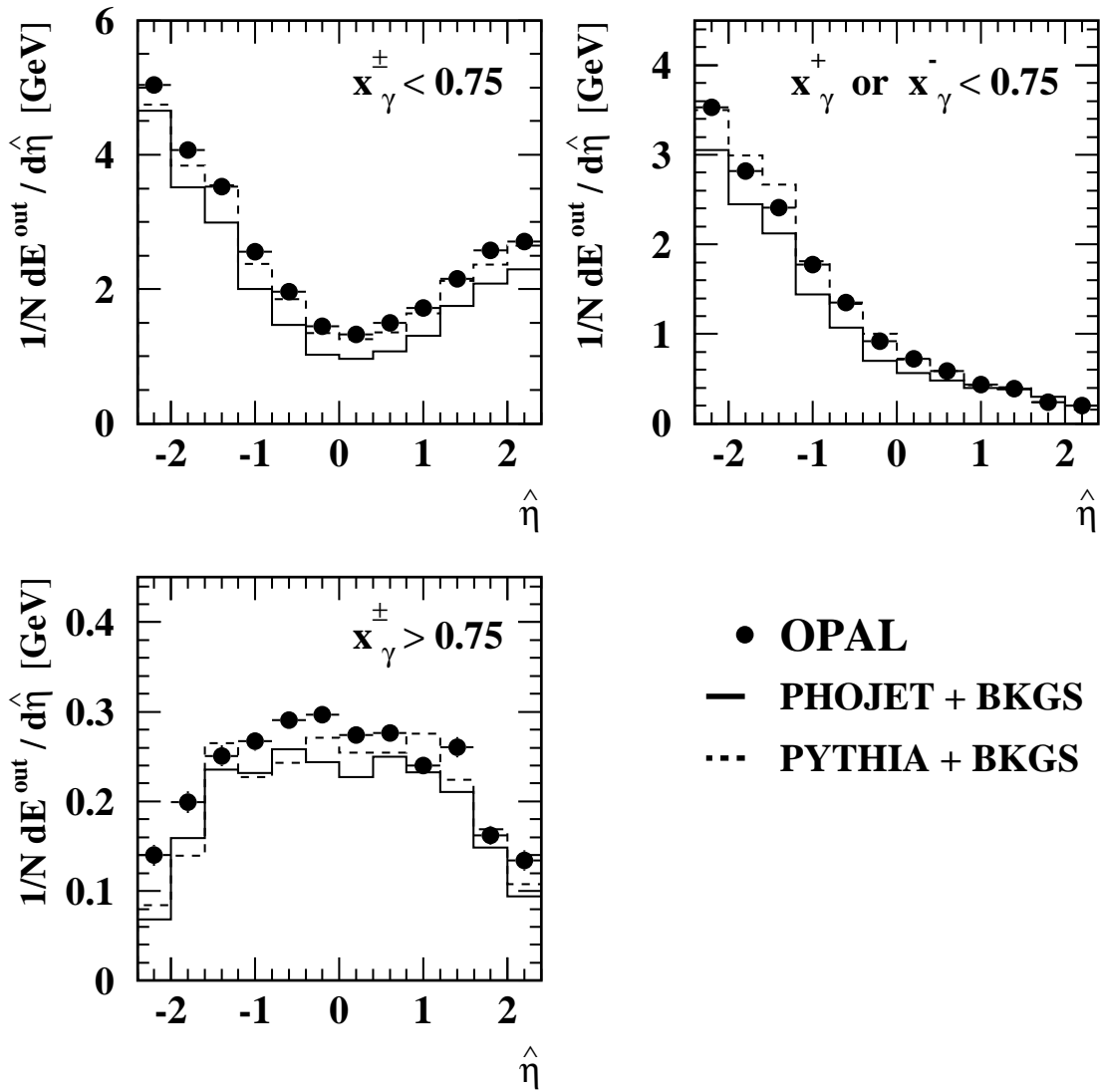}
\end{center}
\caption{ The energy-flow as a function of {\etah} where the area
  around the two leading jets in the event has been excluded in a cone
  of radius 1.3 in $\eta$-$\phi$-space.  The distributions are
  normalised to the number of di-jet events $N$. The statistical error
  is shown where larger than the marker size. The regions in
  {\xgp}-{\xgm}-space considered and the MC simulations used are the
  same as in Figure~\ref{fig:jprophieta}.}
\label{fig:etout}
\end{figure}


\begin{figure}[ht]
\begin{center}
\includegraphics[width=0.95\textwidth]{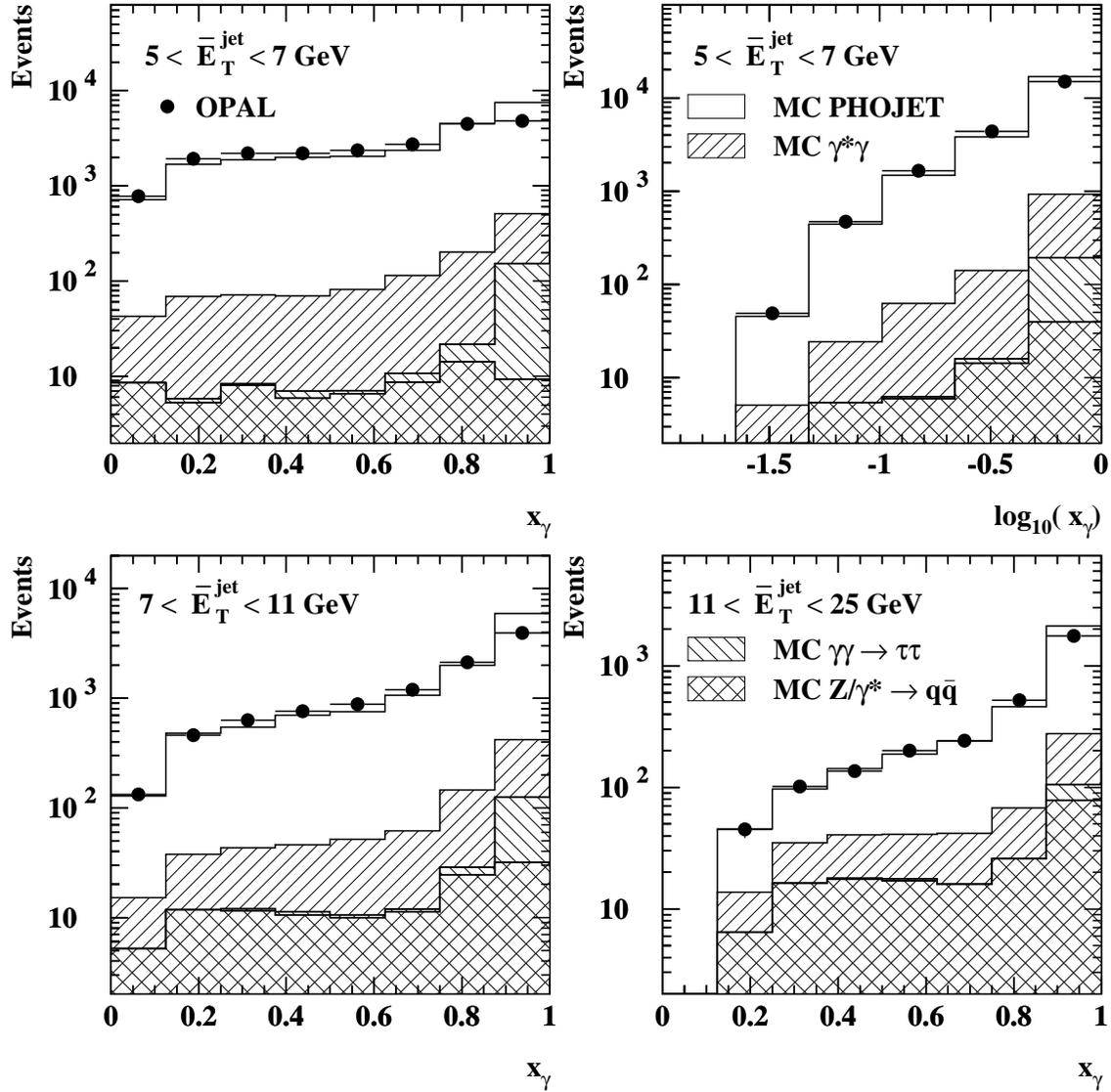}
\end{center}
\caption{  The uncorrected $\xg$ distributions in the data 
  compared to the sum of signal and background processes in the
  simulation. The statistical error is shown where larger
  than the marker size. The MC simulations used are the same as in
  Figure~\ref{fig:selq1}.}
\label{fig:obsdet}
\end{figure}

\cleardoublepage


\begin{figure}[ht]
\begin{center}
\includegraphics[width=0.95\textwidth]{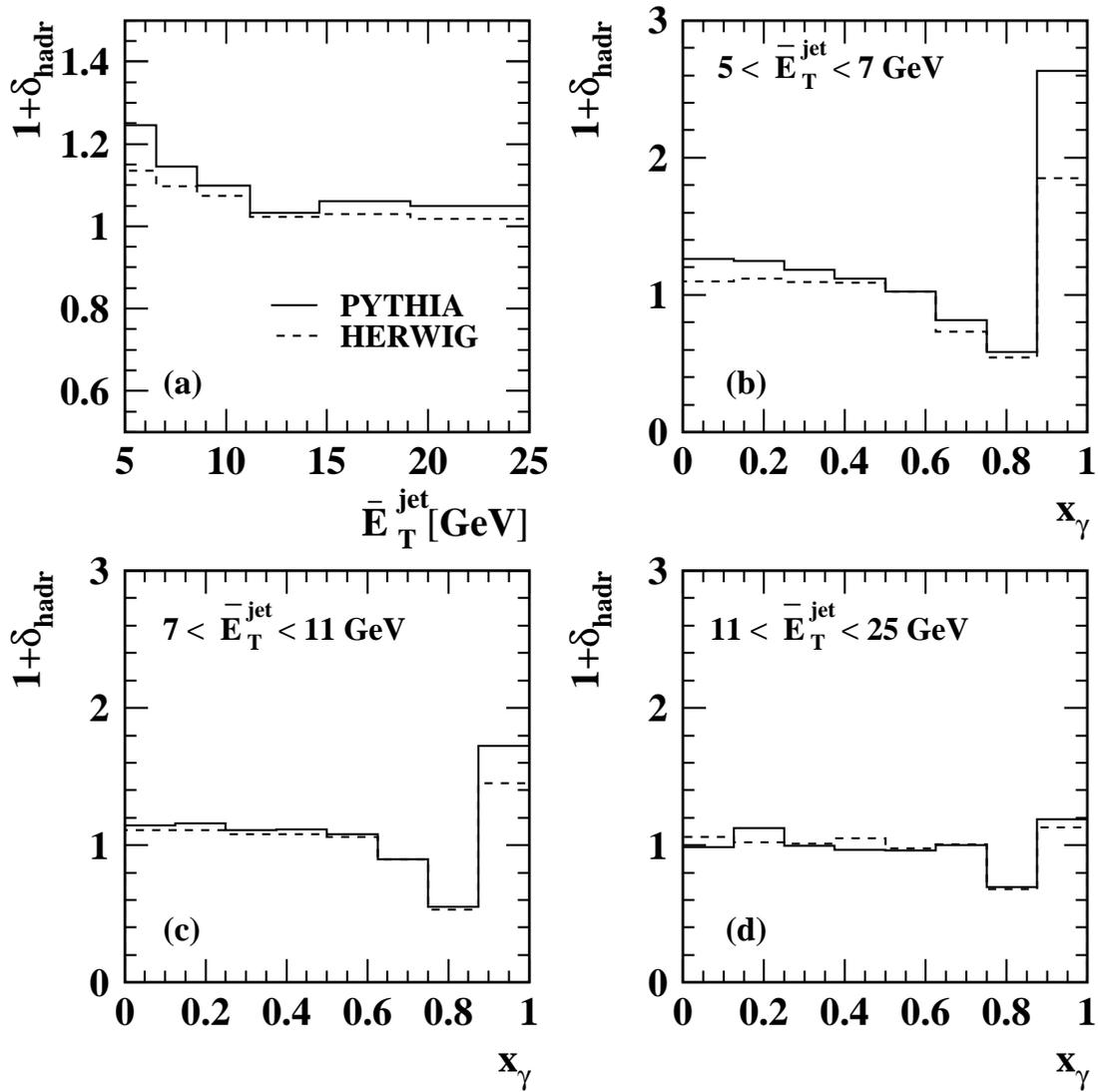}
\end{center}
\caption{ Hadronisation corrections estimated by PYTHIA and HERWIG
  for (a) {\etmean} and (b)-(d) {\xg} for the regions of {\etmean}
  given in the figure. In all cases the full {\xgp}-{\xgm}-space is
  considered. }
\label{fig:hc}
\end{figure}

\cleardoublepage


\begin{figure}[ht]
\begin{center}
\includegraphics[width=0.95\textwidth]{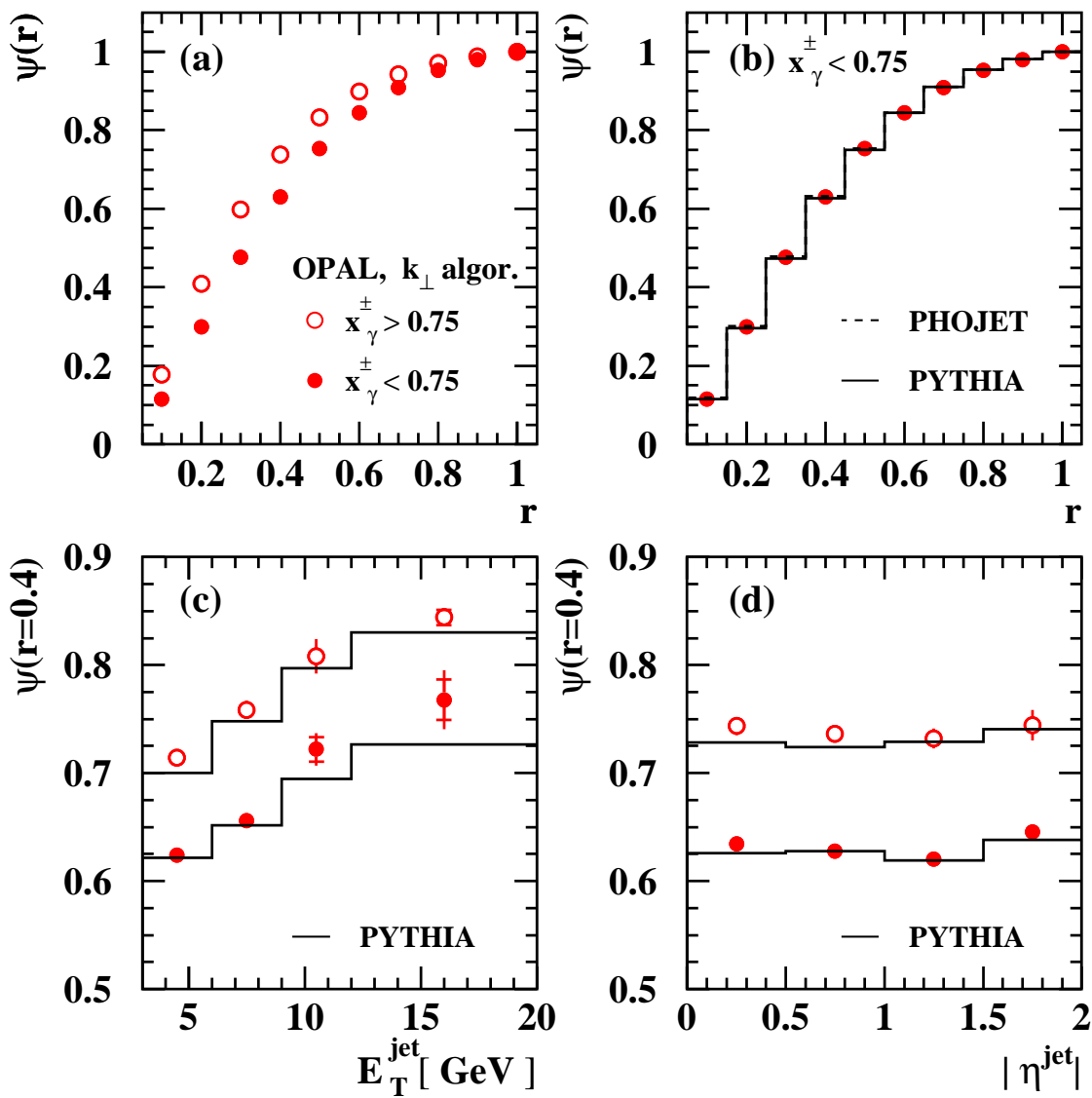}
\end{center}
\caption{The jet shape, $\Psi(r)$, for the two regions of 
  {\xgp}-{\xgm}-space indicated in the figure (a), and $\Psi(r)$ for
  $\xgpm < 0.75$ compared to the predictions of the LO MC generators
  PHOJET and PYTHIA (b). Figures (c) and (d) show the value of
  $\Psi(r=0.4)$ as a function of the transverse energy and
  pseudo-rapidity of the jet respectively, compared to the PYTHIA
  prediction. The total of statistical and systematic uncertainties
  added in quadrature is shown where larger than the marker size. The
  inner error bars show the statistical errors. }
\label{fig:jsh01}
\end{figure}


\begin{figure}[ht]
\begin{center}
\includegraphics[width=0.95\textwidth]{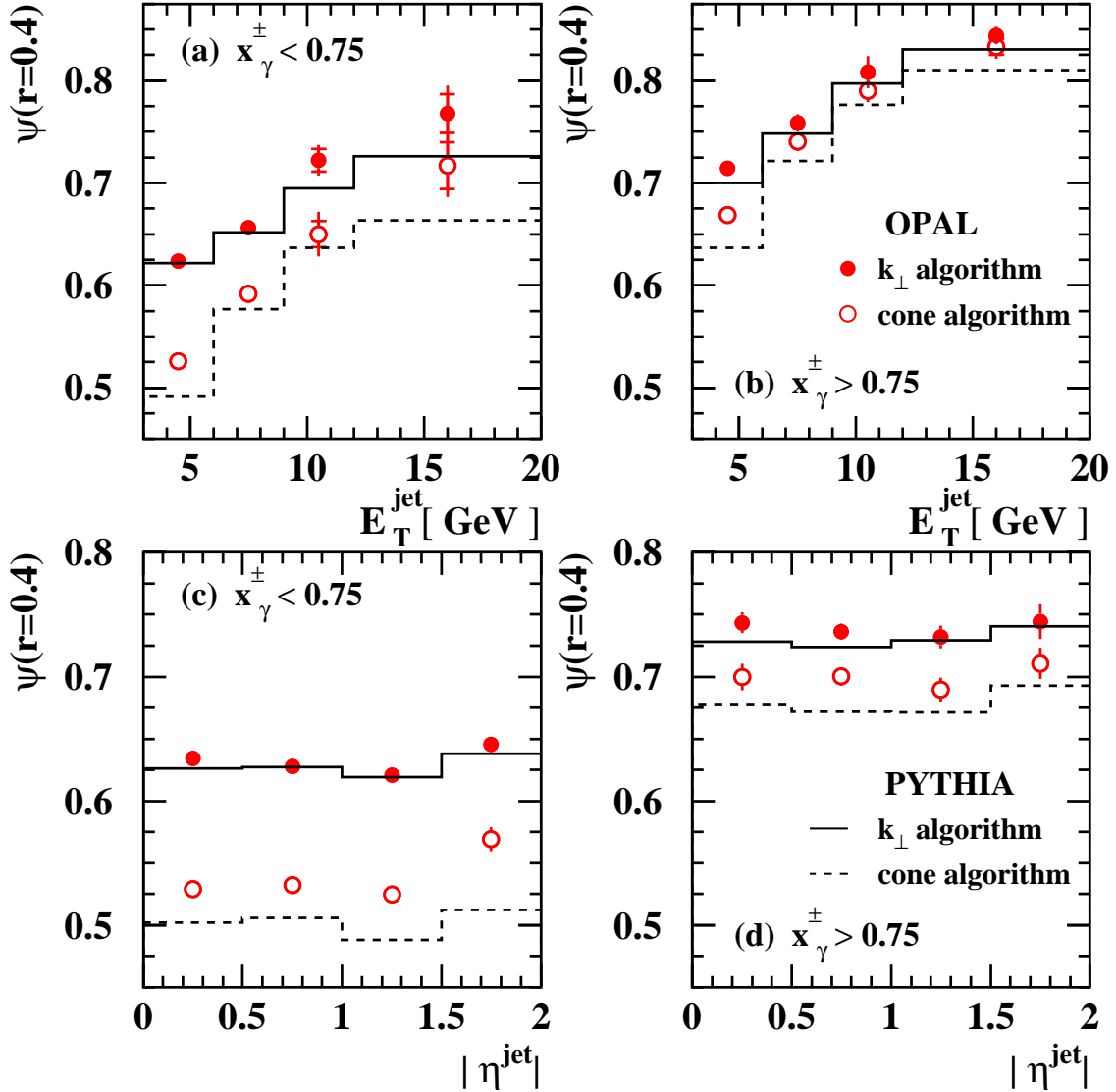}
\end{center}
\caption{The value of the jet shape $\Psi(r)$ at $r=0.4$ as a
  function of the jet transverse energy for ${\xgpm} < 0.75$ (a) and
  ${\xgpm} > 0.75$ (b), and as a function of the jet pseudo-rapidity
  for ${\xgpm} < 0.75$ (c) and ${\xgpm} > 0.75$ (d). In each figure
  the results obtained using the inclusive {\kp} and the cone jet
  algorithm are shown and compared to the PYTHIA prediction. The
  total of statistical and systematic uncertainties added in
  quadrature is shown where larger than the marker size. The inner
  error bars show the statistical errors.}
\label{fig:jsh02}
\end{figure}


\begin{figure}[ht]
\begin{center}
\includegraphics[width=0.95\textwidth]{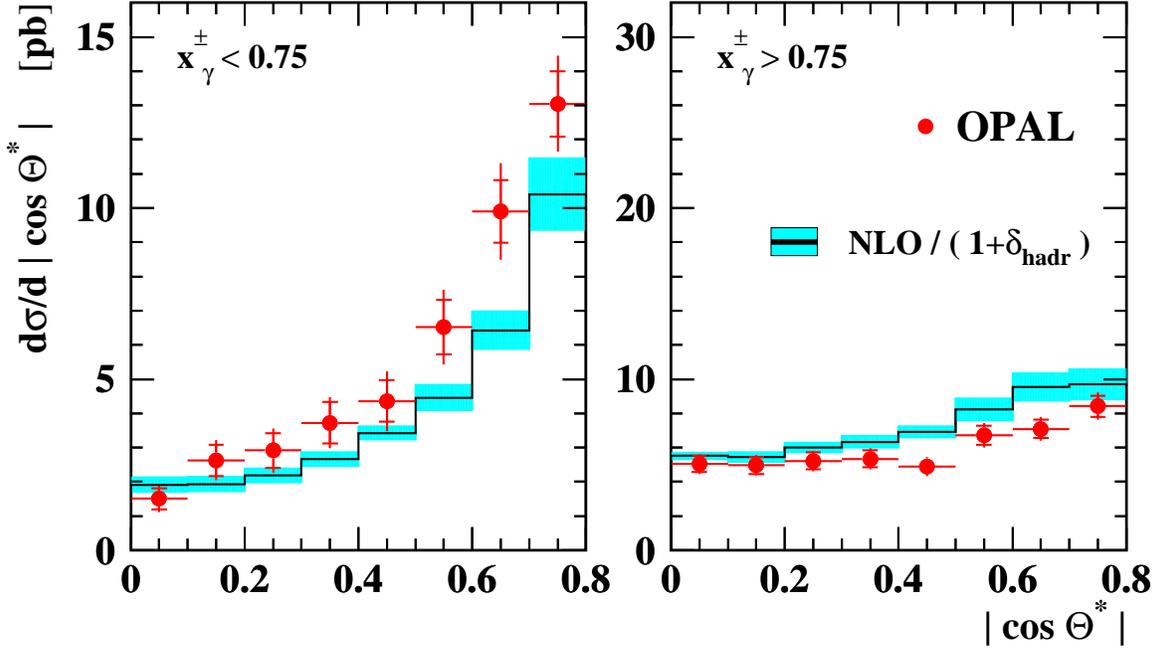}
\end{center}
\caption{The di-jet cross-section as a function of {\costhst} for
  the two regions in {\xgp}-{\xgm}-space indicated in the figure. The
  total of statistical and systematic uncertainties added in
  quadrature is shown where larger than the marker size. The inner
  error bars show the statistical errors. The numerical values are
  given in Table~{\ref{tab:costhst}}. A perturbative NLO QCD
  prediction~\cite{bib-ggnlo} using the {\grvnlo} parton densities is
  compared to the data after hadronisation corrections have been
  applied to the calculation. The shaded band indicates the
  theoretical uncertainty estimated by the quadratic sum of two
  contributions: variation of the renormalisation scale by factors of
  0.5 and 2 and the difference between using HERWIG or PYTHIA in
  estimating the hadronisation corrections.}
\label{fig:costhst}
\end{figure}


\begin{figure}[ht]
\begin{center}
\includegraphics[width=0.95\textwidth]{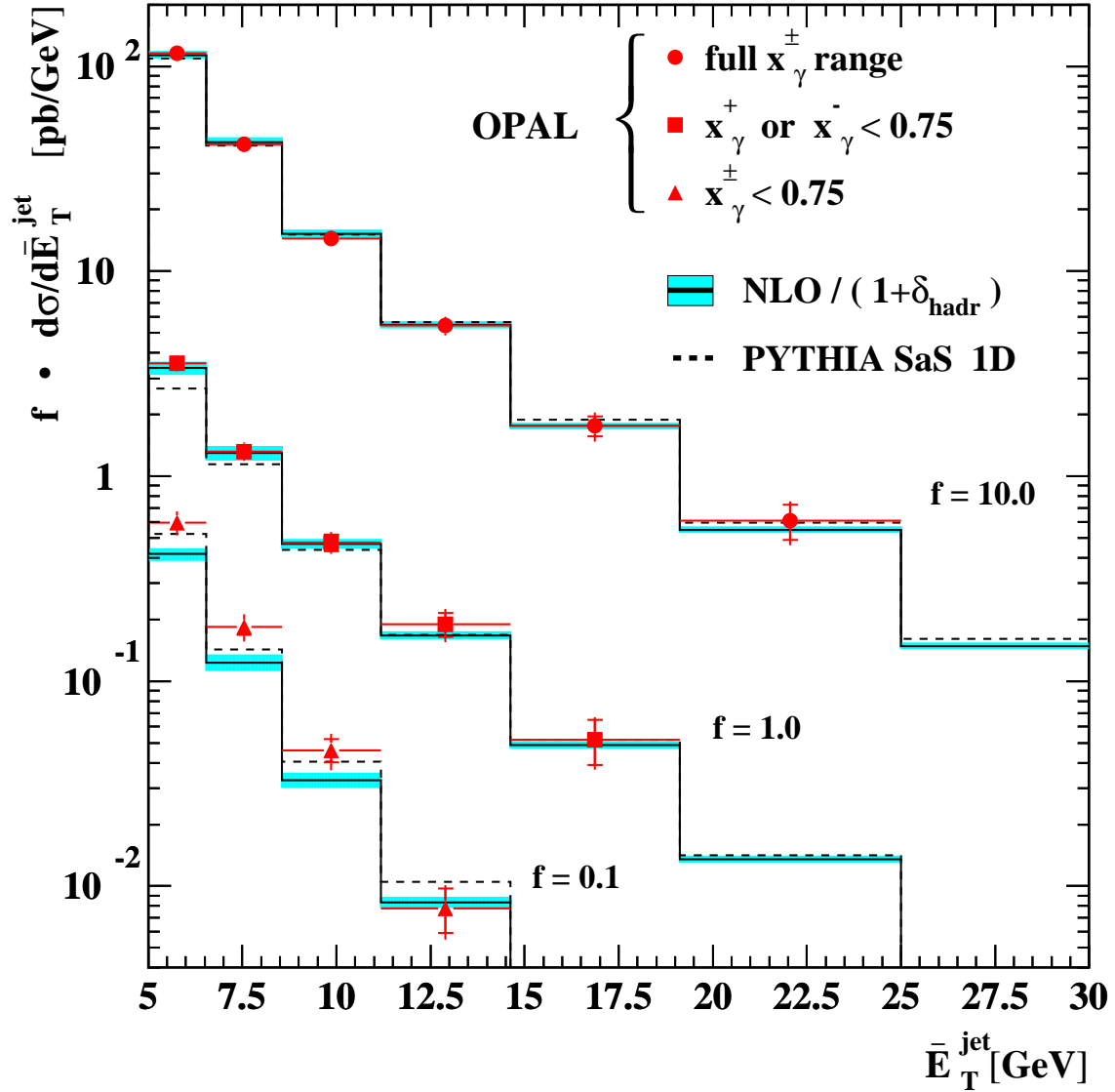}
\end{center}
\caption{The di-jet cross-section as a function of the mean
  transverse energy $\etmean$ of the di-jet system, for the three
  regions in {\xgp}-{\xgm}-space given in the figure.
  The factor $f$ is used to separate the three measurements in the
  figure more clearly. The total of statistical and systematic
  uncertainties added in quadrature is shown where larger than the
  marker size. The inner error bars show the statistical errors. The
  numerical values are given in Table~{\ref{tab:etmxs}}. The
  prediction of the LO program PYTHIA using the parton distribution
  function {\sas1d} is compared to the data. The NLO calculation is
  the same as in Figure~{\ref{fig:costhst}}.}
\label{fig:etmxs}
\end{figure}


\begin{figure}[ht]
\begin{center}
\includegraphics[width=0.95\textwidth]{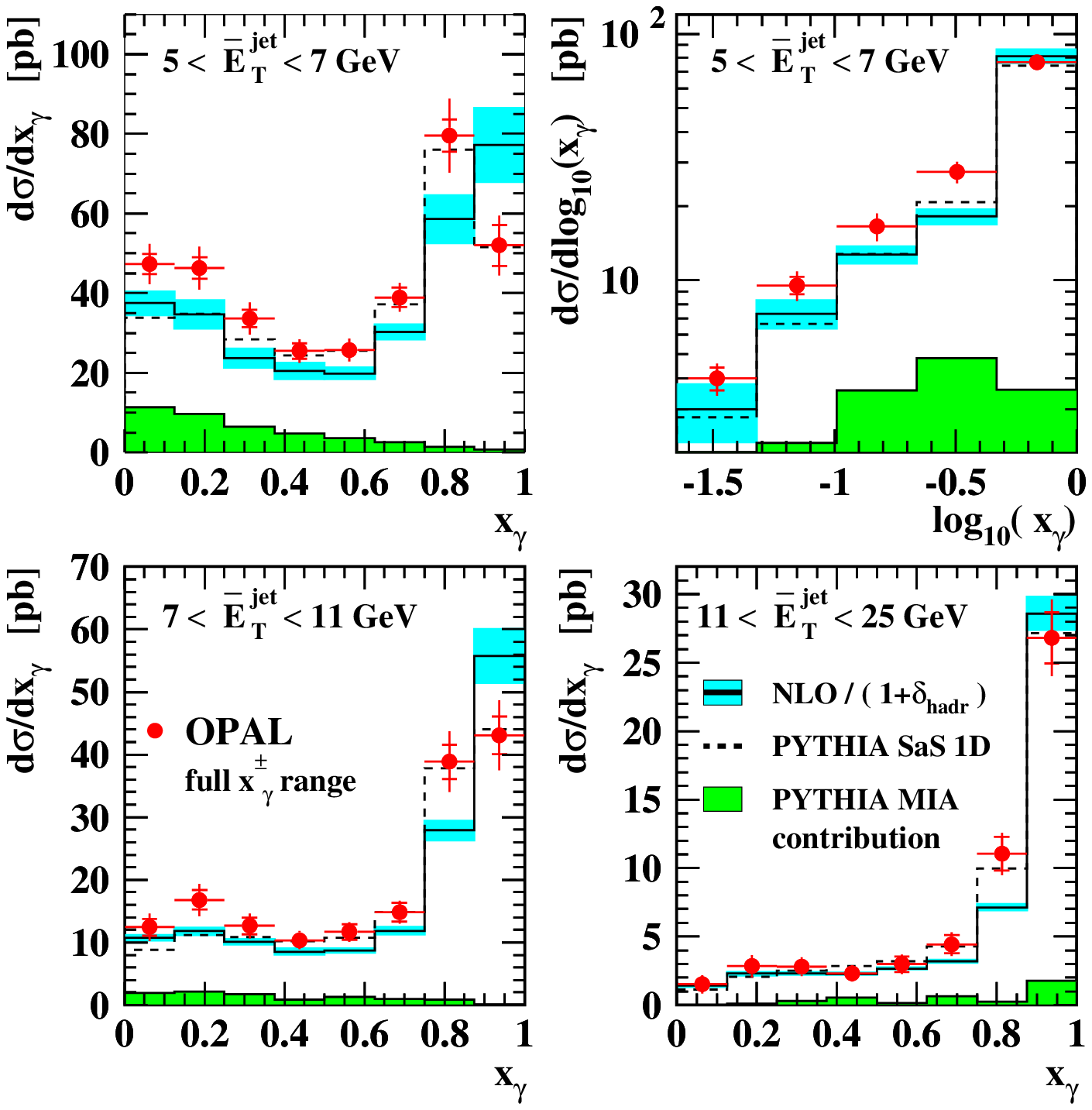}
\end{center}
\caption{The di-jet cross-section as a function of $\xg$ and
  $\mathrm{log_{10}}\left(\xg\right)$ for the regions of the mean
  transverse energy $\etmean$ of the di-jet system indicated in the
  figures. The total of statistical and systematic uncertainties added
  in quadrature is shown where larger than the marker size.  The inner
  error bars show the statistical errors. The numerical values are
  given in Tables~{\ref{tab:xgxs}} and~{\ref{tab:xgxslg}}. The
  prediction of the LO MC generator PYTHIA using the parton distribution
  function {\sas1d} is compared to the data. The shaded histogram at
  the bottom of each plot indicates the MIA contribution to the PYTHIA
  prediction. The NLO calculation is the same as in
  Figure~{\ref{fig:costhst}}.}
\label{fig:xgxs}
\end{figure}

\begin{figure}[ht]
\begin{center}
\includegraphics[width=0.95\textwidth]{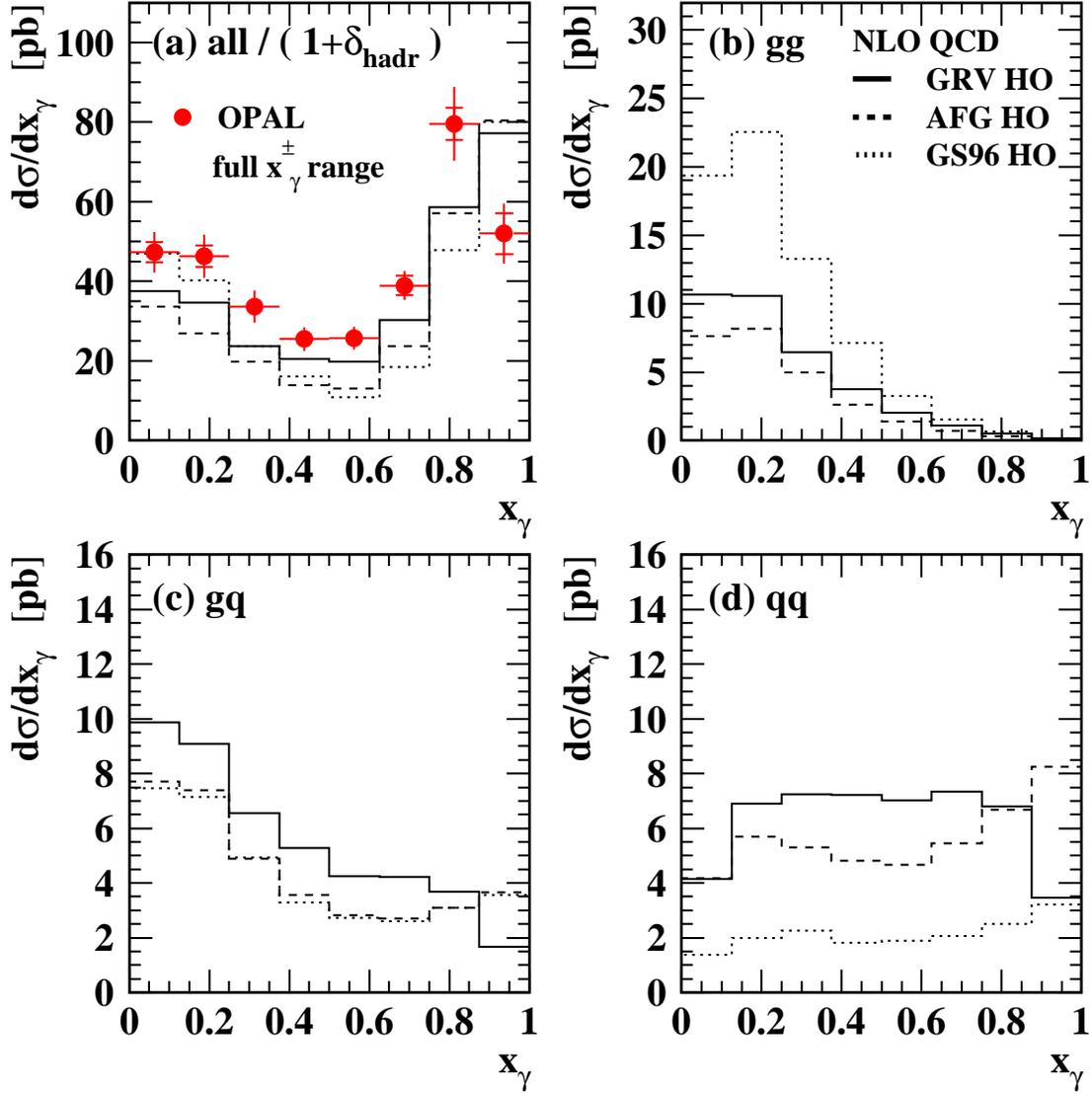}
\end{center}
\caption{The prediction of NLO QCD using different 
  parton distributions for the photons. The di-jet cross-section as a
  function of $\xg$ for 5~GeV $<$ {\etmean} $<$ 7~GeV (upper left plot
  in Figure~\ref{fig:xgxs}) is shown. In (a) the full cross-section is
  shown after hadronisation corrections have been applied, while (b),
  (c) and (d) show the gg, gq and qq contributions to this
  cross-section without hadronisation corrections. The NLO calculation
  is the same as in Figure~{\ref{fig:costhst}}. }
\label{fig:nlocont}
\end{figure}

\begin{figure}[ht]
\begin{center}
\includegraphics[width=0.95\textwidth]{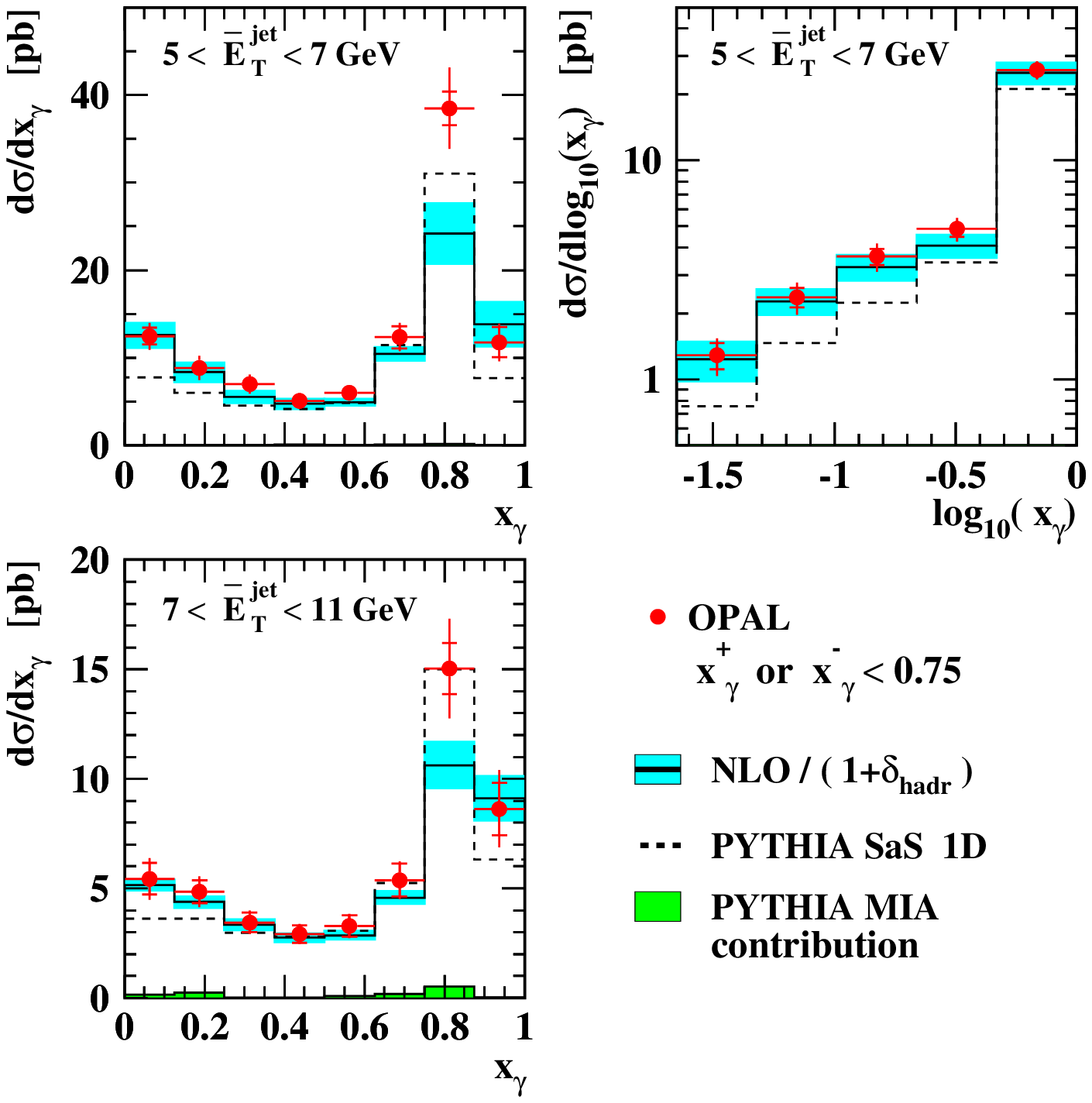}
\end{center}
\caption{The di-jet cross-section as a function of $\xg$ and
  $\mathrm{log_{10}}\left(\xg\right)$ for the regions of the mean
  transverse energy $\etmean$ of the di-jet system indicated in the
  figures and for ${\xgp}$ or ${\xgm} < 0.75$ . The total of
  statistical and systematic uncertainties added in quadrature is
  shown where larger than the marker size.  The inner error bars show
  the statistical errors. The numerical values are given in
  Tables~{\ref{tab:xgxssr}} and~{\ref{tab:xgxslg}}. The NLO
  calculation and MC simulation are the same as in
  Figure~{\ref{fig:xgxs}}.}
\label{fig:xgxssr}
\end{figure}


\begin{figure}[ht]
\begin{center}
\includegraphics[width=0.95\textwidth]{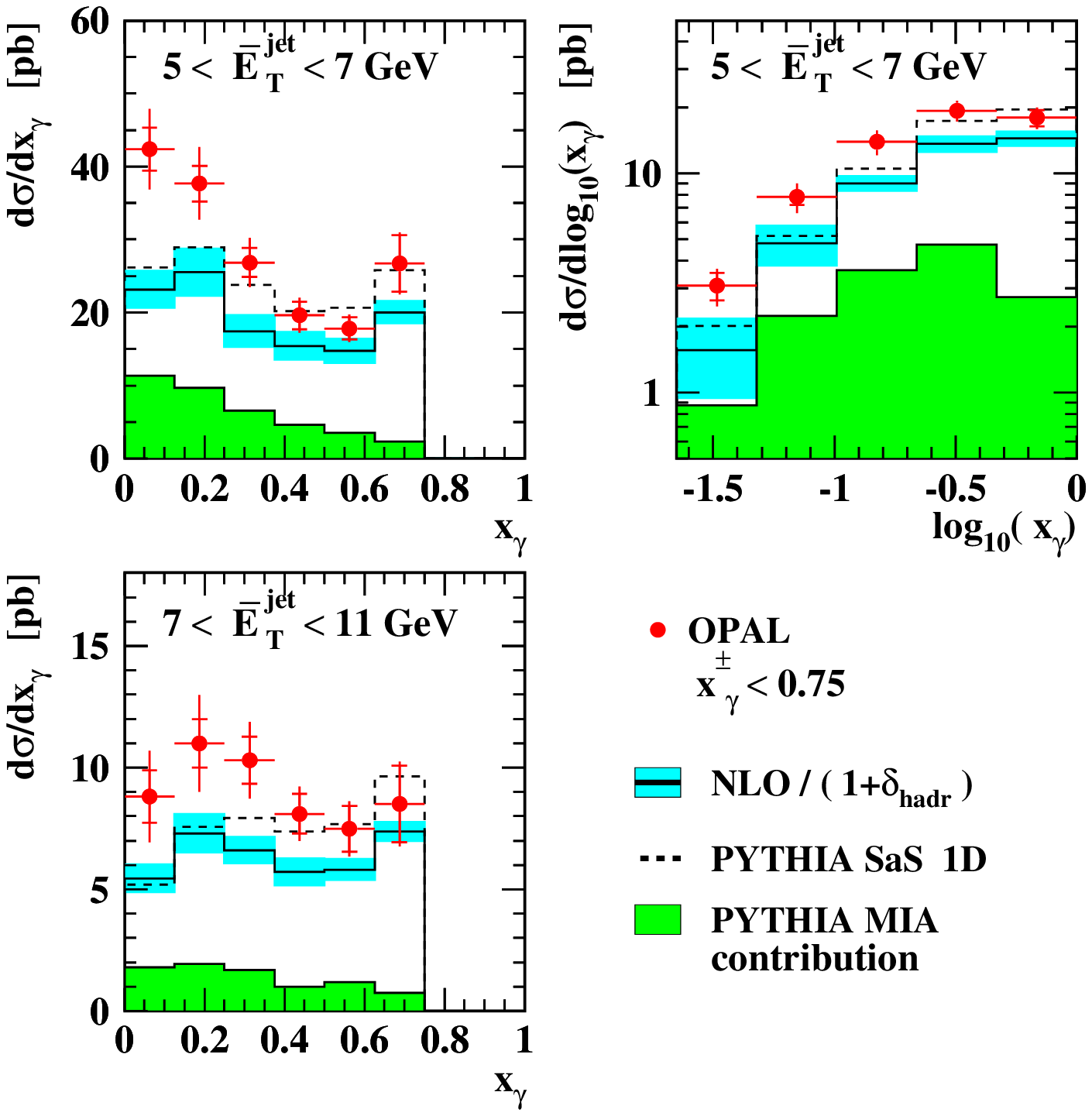}
\end{center}
\caption{The di-jet cross-section as a function of $\xg$ and
  $\mathrm{log_{10}}\left(\xg\right)$ for the regions of the mean
  transverse energy $\etmean$ of the di-jet system indicated in the
  figures.  For these cross-sections ${\xgpm} < 0.75$ is required. The
  total of statistical and systematic uncertainties added in
  quadrature is shown where larger than the marker size. The inner
  error bars show the statistical errors.  The numerical values are
  given in Tables~{\ref{tab:xgxsdr}} and~{\ref{tab:xgxslg}}. The NLO
  calculation and MC simulation are the same as in
  Figure~{\ref{fig:xgxs}}.}
\label{fig:xgxsdr}
\end{figure}


\begin{figure}[ht]
\begin{center}
\includegraphics[width=0.95\textwidth]{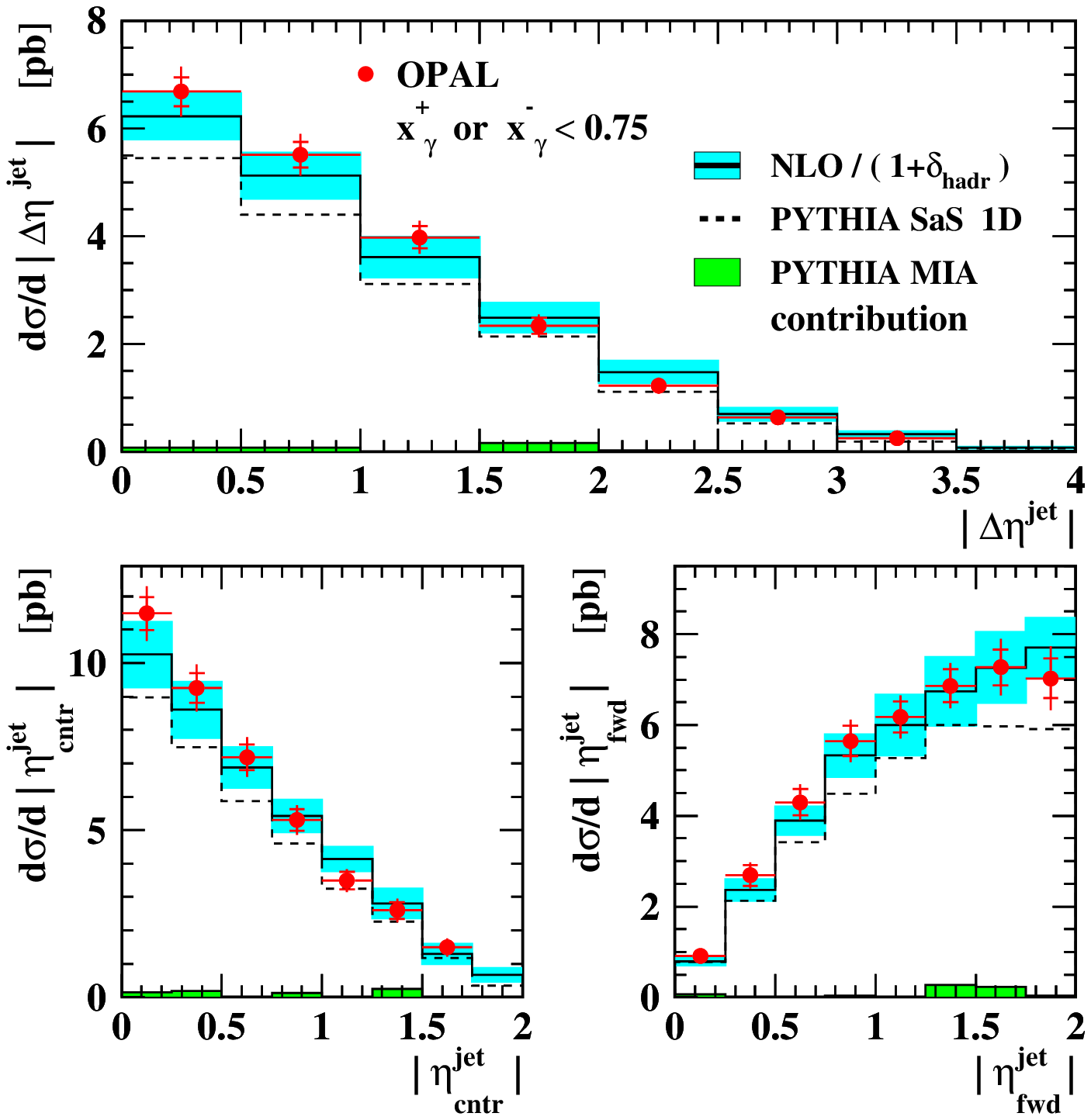}
\end{center}
\caption{The di-jet cross-section as a function of {\detajet} for
  the two leading jets in {\etjet} and separately for the central and
  the forward jet, for ${\xgp}$ or ${\xgm} < 0.75$. The total of
  statistical and systematic uncertainties added in quadrature is
  shown where larger than the marker size. The inner error bars show
  the statistical errors. The numerical values are given in
  Tables~{\ref{tab:d-etaxs}},~{\ref{tab:c-etaxs}}
  and~{\ref{tab:f-etaxs}}. The NLO calculation and MC simulation are the
  same as in Figure~{\ref{fig:xgxs}}.}
\label{fig:etaxssr}
\end{figure}


\begin{figure}[ht]
\begin{center}
\includegraphics[width=0.95\textwidth]{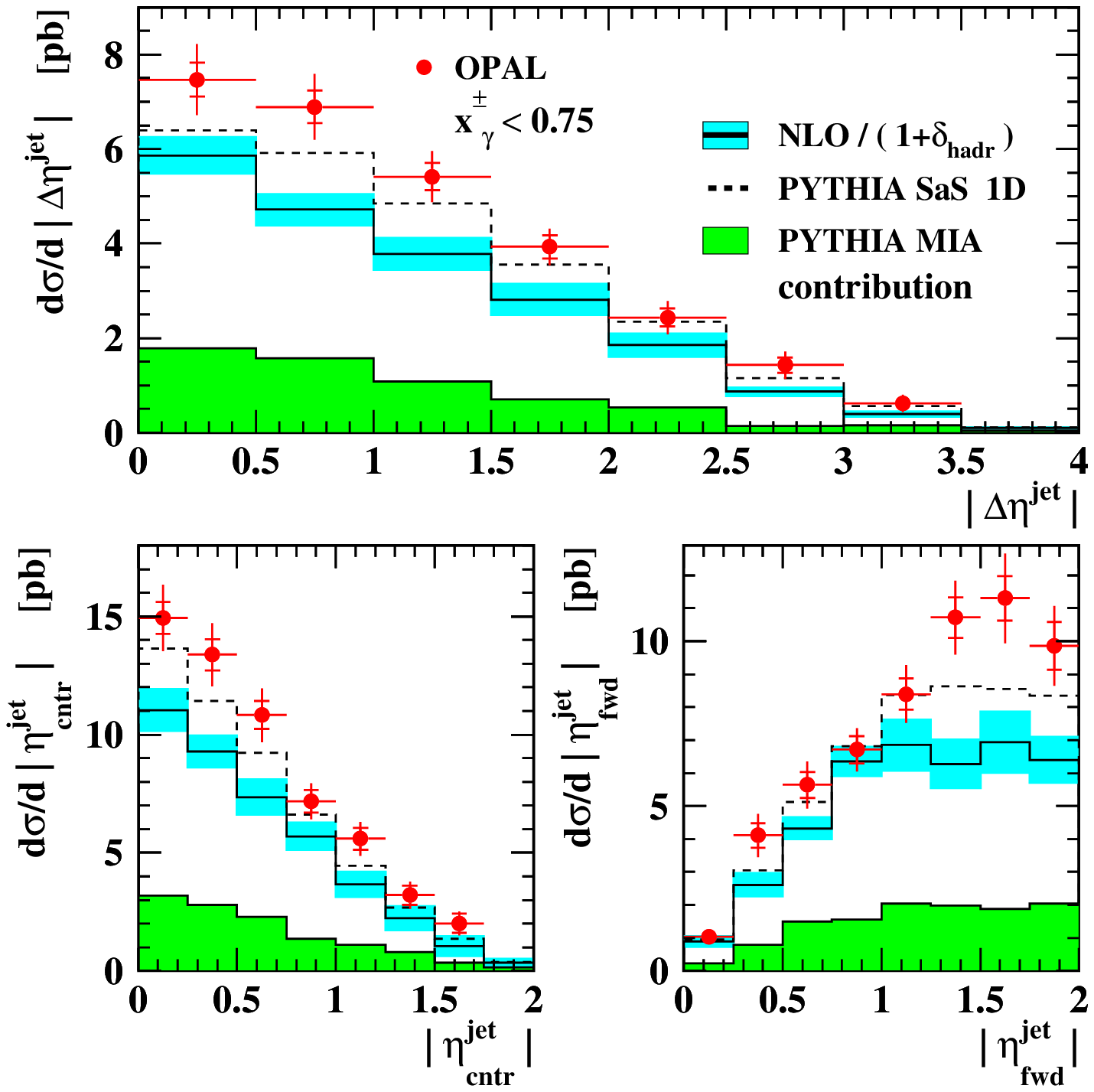}
\end{center}
\caption{The di-jet cross-section as a function of {\detajet} for
  the two leading jets in {\etjet} and separately for the central and
  the forward jet.  For these cross-sections ${\xgpm} < 0.75$ is
  required. The total of statistical and systematic uncertainties
  added in quadrature is shown where larger than the marker size. The
  inner error bars show the statistical errors. The numerical values
  are given in Tables~{\ref{tab:d-etaxs}},~{\ref{tab:c-etaxs}}
  and~{\ref{tab:f-etaxs}}. The NLO calculation and MC simulation are
  the same as in Figure~{\ref{fig:xgxs}}.}
\label{fig:etaxsdr}
\end{figure}

\end{document}